\documentclass[10pt,oneside,twocolum,final]{IEEEtran}
\usepackage{amsfonts}
\usepackage{amsmath}
\usepackage{graphicx}
\usepackage{color}
\usepackage{multicol}
\usepackage{amssymb}
\usepackage{subfigure}
\usepackage{bm}
\usepackage{latexsym}
\usepackage{stfloats}
\usepackage{float}
\usepackage{exscale}
\usepackage{relsize}
\usepackage{cite}
\usepackage{cases}
\usepackage{algorithm}
\usepackage{algpseudocode}
\usepackage{amsmath}
\usepackage{graphics}
\usepackage{graphicx}
\usepackage{epsfig}
\usepackage{epstopdf}
\usepackage{breqn}
\usepackage{enumerate}
\usepackage{multirow}
\usepackage[utf8]{inputenc}

\begin{document}
	
	\title{Multiuser Full-Duplex Two-Way Communications via Intelligent Reflecting Surface}
	\IEEEoverridecommandlockouts
	\author{
		Zhangjie~Peng,
		Zhenkun~Zhang,
		Cunhua Pan,
		Li~Li,
		and~A. Lee Swindlehurst,~\IEEEmembership{Fellow,~IEEE}
		\thanks{Z. Peng is with the College of Information, Mechanical and Electrical Engineering,
			Shanghai Normal University, Shanghai 200234, China,
			and also with the National Mobile Communications Research Laboratory, Southeast University,
			Nanjing 210096, China (e-mail: pengzhangjie@shnu.edu.cn).
			L. Li and Z. Zhang are with the College of Information, Mechanical and Electrical Engineering,
			Shanghai Normal University, Shanghai 200234, China (e-mails: lilyxuan@shnu.edu.cn, 1000479070@smail.shnu.edu.cn).
			C. Pan is with the School of Electronic Engineering and Computer Science at Queen Mary University of London,
			London E1 4NS, U.K.  (e-mail: c.pan@qmul.ac.uk).
			A. L. Swindlehurst is with the Department of Electrical Engineering and Computer Science,
			University of California at Irvine, Irvine, CA 92697 USA (e-mail: swindle@uci.edu).
			The work of Z. Peng was supported in part by the NSFC under 61701307,
			and the open research fund of National Mobile Communications Research Laboratory, Southeast University (No. 2018D14);
			the work of A. L. Swindlehurst was supported in part by the U.S. National Science Foundation under grant ECCS-2030029.
			\emph{ (Corresponding author: Cunhua Pan and Zhenkun Zhang)}}
	}
	
	
	

	\maketitle
	
	\newtheorem{lemma}{Lemma}
	\newtheorem{theorem}{Theorem}
	\newtheorem{remark}{Remark}
	\newtheorem{corollary}{Corollary}
	\newtheorem{proposition}{Proposition}

	\begin{abstract}
		Low-cost passive intelligent reflecting surfaces (IRSs) have recently been envisioned as a revolutionary technology capable of reconfiguring the wireless propagation environment through carefully tuning reflection elements.
		This paper proposes deploying an IRS to cover the dead zone of cellular multiuser full-duplex (FD) two-way communication links while suppressing user-side self-interference (SI) and co-channel interference (CI).
		This approach, allowing the base station (BS) and all users to exchange information simultaneously, can potentially double the spectral efficiency.
		To ensure network fairness, we jointly optimize the precoding matrix of the BS and the reflection coefficients of the IRS to maximize the weighted minimum rate (WMR) of all users, subject to maximum transmit power and unit-modulus constraints.
		We reformulate this non-convex problem and decouple it into two subproblems.
		Then the optimization variables in the equivalent problem are alternately optimized by adopting the block coordinate descent (BCD) algorithm.
		In order to further reduce the computational complexity, we propose the minorization-maximization (MM) algorithm for optimizing the precoding matrix and the reflection coefficient vector by defining minorizing functions in the surrogate problems.
		Finally, simulation results confirm the convergence and efficiency of our proposed algorithm, and validate the advantages of introducing IRS to improve coverage in blind areas.
		
		
	\end{abstract}

	\section{Introduction}
	In the future 5G-and-beyond era, wireless networks will be required to achieve a 1000-fold increase in capacity compared with current networks, motivated by the growing popularity of applications that rely on high data rate transmission, such as three-dimensional (3D) video and augmented reality (AR) \cite{itu2015vision}.
	To achieve this progress, promising techniques such as millimeter wave (mmWave) communication, ultra-dense cloud radio access networks (UD-CRAN) \cite{8387197} and massive multiple-input multiple-output (M-MIMO) arrays \cite{7880689} have been advocated \cite{6824752}.
	In addition, full-duplex (FD) two-way communication in which two or more devices simultaneously exchange data at the same carrier frequency has received extensive research attention as it can double the spectral-efficiency of the wireless communication system \cite{6403869,6690332}.
	Due to its appealing advantages, two-way FD relaying has been extensively studied in various scenarios, such as D2D communications \cite{6403869}, cognitive radio \cite{7925681}, mmWave communication \cite{gong2017millimeter-wave} and M-MIMO \cite{zhang2016spectral}.
	However, an FD two-way network suffers from low energy-efficiency and high hardware cost.
	For example, the large number of antennas in M-MIMO leads to a large number of RF chains and incurs high power consumption, while energy-intensive transceivers and complex signal processing techniques are required to support the mmWave communication.
	Moreover, another non-negligible bottleneck in the implementation of FD two-way communications lies in the propagation environment.
	In particular, besides the loop-interference (LI) at the relay, this network must also overcome back-propagation interference at the base station (BS) and the users.
	
	Thanks to breakthroughs in micro-electrical-mechanical systems and programmable metamaterials, the intelligent reflecting surfaces (IRSs) have recently attracted extensive attention from researchers as a means to improve both the spectral- and energy-efficiency of wireless communications networks \cite{di2019smart},
	and to enable the future vision of smart radio environments \cite{9140329}.
	An IRS comprises a number of low-cost passive reflection elements requiring no dedicated energy sources \cite{9136592},
	and each reflection element can independently impose a continuously or discretely tunable phase shift onto the incident signal \cite{cui2014coding,liu2019intelligent}.
	When the phase shifts are properly adjusted, the directly transmitted signal and the reflected signal can be superimposed constructively at the intended receivers or destructively at other unintended users.
	Note that an IRS can also implement fine-grained 3D passive beamforming \cite{wu2019towards}, and thus its function resembles that of an FD MIMO amplify-and-forward (AF) relay.
	The difference is that the IRS transmits signals through passive reflection, requiring no signal processing to deal with LI and leading to negligible energy consumption.
	In addition, unlike active relay transmission, an IRS does not generate new signals or thermal noise.
	Thanks to its miniaturized circuits, an IRS also has the attractive advantages of light weight, small size and high integration, which enables it to be used to improve  indoor propagation environments \cite{8485924}.
	For outdoor communication scenarios, it can be integrated into the existing infrastructure, such as building facades, station signs and lampposts.
	
	Due to these promising features, joint precoding at the BS/AP and reflecting at the IRS has been extensively studied in one-way communication networks, for the
	MISO case \cite{nadeem2019asymptotic,xu2019resource,2020arXiv200511663H,8741198},
	physical layer security \cite{xu2019resource,2019arXiv190409573Y,9133130}, simultaneous wireless information and power transfer (SWIPT) \cite{pan2019intelligent}, mobile edge computing \cite{bai2019latency}, and multigroup multicast \cite{zhou2019intelligent}.
	In addition, the deep reinforcement learning technique has been leveraged for this joint design \cite{9110869}.
	More system factors, such as channel estimation and the overhead required for configuring the phase shifts, are taken into account in recent works \cite{9200578}.
	However, there is a paucity of investigations on the study of the integration of IRS in two-way communications \cite{2020arXiv200107907A,9025235,xu2020resource}.
	The work of \cite{2020arXiv200107907A} and \cite{9025235} considered communication between two SISO end users and two MIMO sources, respectively,
	both of which are aimed at maximizing the system sum rate.
	A cognitive radio system consisting of an FD BS and multiple half-duplex users was considered in \cite{xu2020resource},
	where the system sum rate of the secondary network was maximized with a constraint on the interference to the primary users.
	However, the fairness between uplink and downlink transmissions needs to be guaranteed in FD communication, and this has not been taken into account in these studies.
	
\begin{figure}
		\centering
		\includegraphics[scale=0.67]{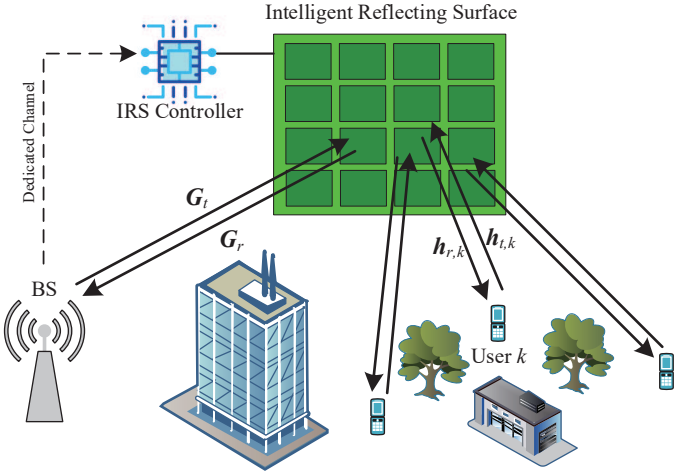}
		\caption{Illustration of the IRS-aided FD two-way communication between a MIMO BS and $K$ SISO users.}
		\label{Figsysmodel}
	\end{figure}

	In this paper, we propose to employ an IRS in an FD two-way network to provide signal coverage for users in blind areas, as shown in Fig. \ref{Figsysmodel}.
	Specifically, unlike the relay schemes in \cite{zhang2016two-timeslot}, in our proposed system, both the uplink and downlink transmissions can occur simultaneously and operate at the same frequency via the reflection of the IRS, and thus potentially doubles the spectral-efficiency.
	In order to guarantee fairness, the max-min fairness (MMF) criterion is chosen as the optimization metric, which is a complex non-differentiable objective function (OF) that cannot be solved by applying the existing methods proposed in the related works such as \cite{nadeem2019asymptotic}.
	
	We summarize the main contributions and challenges of this work as follows
	
	\begin{enumerate}
		\item
		To the best of our knowledge, this is the first work to consider fairness in a multiuser FD two-way communication network with the assistance of an IRS.
		Specifically, we jointly optimize the precoding matrix of the BS and the reflection coefficients of the IRS to maximize the weighted minimum rate (WMR) of all users, subject to maximum transmit power and unit modulus constraints.
		This problem is challenging to tackle for the non-differentiable \mbox{OF and the highly coupled optimization variables.}
		\item
		By applying the weighted minimum mean-square error (WMMSE) criterion and introducing certain auxiliary variables, the original problem is transformed and solved effectively through the proposed block coordinate descent (BCD) algorithm, in which each set of variables is alternately optimized.
		In particular, the precoding subproblem is formulated as a second-order cone programming problem (SOCP), and the reflection coefficient subproblem is derived as a quasi-SOCP with a non-convex quadratic constraint.
		\item
		In order to further reduce the computational complexity of the BCD algorithm, we proposed a modified Minorization-Maximization (MM) algorithm.
		Specifically, unlike the quadratic form in \cite{8741198}, the OFs of both subproblems are non-differentiable.
		We thus utilize the smooth approximation theory \cite{Xu2001Smoothing} to obtain differentiable approximations for them.
		Then, the corresponding minorizing functions are derived sequentially, which leads to surrogate problems with closed-form solutions.
		Hence, both approximated subproblems are solved efficiently by the MM algorithm in an iterative manner.
		\item
		Our simulation results illustrate the feasibility of the proposed approach and the advantages of using an IRS in assisting the FD two-way communication.
		Additionally, the results also provide guidance for practical engineering designs, and highlight the trade-off between improved self-interference (SI) elimination when the IRS is deployed near the users, and reduced propagation blockages when the IRS is deployed near the BS.
		The convergence and the efficiency of the proposed algorithm are also verified.
	\end{enumerate}
	
	The rest of the paper is organized as follows.
	Section \ref{SYSTEM MODEL AND PROBLEM FORMULATION} describes the system model involving multiuser FD two-way communication via an IRS, and formulates the WMR maximization problem.
	In Section \ref{SOCP-BASED BCD METHOD}, we derive the subproblems corresponding to each set of variables by reformulating the original problem and performing alternating optimization.
	In Section \ref{LOW-COMPLEXITY ALGORITHM DEVELOPMENT}, we propose a low-complexity version of the algorithm.
	Extensive simulation results are presented in Section \ref{SIMULATION RESULTS}.
	Finally, we conclude the paper in Section \ref{CONCLUSION}.
	
	\emph{Notation}:
	Vectors and matrices are denoted by boldface lower and boldface capital case letters, respectively.
	The quantities ${\bf a}_m$ and ${\bf A}_{m,n}$ respectively denote the $m$th element of vector $\bf a$ and the $\left(m,n\right)$-entry of matrix $\bf A$.
	${{\mathbb C}^{{M} \times N}}$ denotes the space of ${{M} \times N}$ complex-valued matrices, and $j \triangleq \sqrt {{\rm{ - 1}}}$ is the imaginary unit.
	${\bf A}^{\rm H}$, ${\bf A}^{\rm T}$ and ${\bf A}^{*}$ denote the Hermitian, transpose and conjugate of matrix ${\bf A}$, respectively.
	The trace and Frobenius norm of a matrix are denoted by ${\rm Tr}\left[\cdot\right]$ and $\left\| \cdot \right\|_F$, respectively.
	$\left\| \cdot \right\|_1$ and $\left\| \cdot \right\|_2$ denote the $l_1$- and $l_2$-norm of a vector, respectively.
	For a complex scalar $a$, ${\rm Re}\left\{a\right\}$, ${\mathbb E}\left[a\right]$, $\left|a\right|$ and $\angle\left(a\right)$ denote the real part, expectation, absolute value and angle of $a$, respectively.
	The functions ${\rm diag}\left(\cdot\right)$ and ${\rm vec}\left(\cdot\right)$ represent diagonalization and vectorization operators.
	${\bf A} \succeq {\bf B}$ means that ${\bf A} - {\bf B}$ is a positive semidefinite matrix.
	The Hadamard product and Kronecker product of ${\bf A}$ and ${\bf B}$ are respectively denoted by ${\bf A} \odot {\bf B}$ and ${\bf A} \otimes {\bf B}$.
	${\mathcal C}{\mathcal N}(0,\sigma^2)$ denotes \mbox{the Gaussian distribution with mean $0$ and variance $\sigma^2$.}

	\section{System Model and Problem Formulation}\label{SYSTEM MODEL AND PROBLEM FORMULATION}

	\subsection{Signal Transmission Model}
	Consider an FD two-way communication system with one BS and multiple users, where both the downlink and uplink transmissions occur at the same time and the same frequency as shown in Fig. \ref{Figsysmodel}.
	Due to path loss and blockages, no direct link between the BS and the users is assumed to exist.
	An IRS is deployed to assist the data transmission by establishing additional non-line-of-sight (NLoS) links.
	
	The BS is equipped with ${N_{\rm{t}}} > 1$ transmit antennas and ${N_{\rm{r}}} > 1$ receive antennas.
	In the service area of the IRS, there are $K$ users, each equipped with a pair of transmit and receive antennas.
	Additionally, we assume that each user transmits signals with a fixed power.
	
	The signal transmitted from the BS is given by
	\begin{equation}
		{\bf{x}}_{\rm{D}} = \sum\limits_{k = 1}^K {{{\bf{f}}_k}{s_{{\rm{D}},k}}},
	\end{equation}
	where ${{s_{{\rm{D}},k}}}$ denotes the desired data symbol for user $k$ and ${{\mathbf{f}}_k} \in {{\mathbb C}^{{N_{\rm t}} \times 1}}$ is the corresponding beamforming vector.
	Similarly, the transmit signal at user $k$ is
	\begin{equation}
		{x_{{\rm{U}},k}} = \sqrt {{P_k}} {s_{{\rm{U}},k}},
	\end{equation}
	where $s_{{\rm{U}},k}$ denotes the data symbol sent by user $k$, and $P_k$ is the corresponding transmit power.
	Defining ${\mathcal L}=\left\{ {{\rm{D}}, {\rm{U}}} \right\}$ and ${\mathcal K}=\left\{ {1, \cdots ,K} \right\}$, we assume each $s_{l,k}$ for $\forall l \in {\mathcal L},k \in {\mathcal K}$ is an independent Gaussian data symbol and has unit power,
	i.e., ${\mathbb E}\left[ {{s_{l,k}}s_{l,k}^\ast} \right] = 1$ and ${\mathbb E}\left[ {{s_{l,k}}s_{i,j}^\ast} \right] = 0$, $\left\{ {l,k} \right\} \ne \left\{ {i,j} \right\}$.
	Let us denote ${\mathbf{F}} = \left[ {{{\mathbf{f}}_1}, \cdots ,{{\mathbf{f}}_K}} \right] \in {{\mathbb C}^{{N_t} \times K}}$ as the collection of all beamforming vectors, so that the power constraint of the BS can be written as
	\begin{equation}\label{pow_constraint}
		{{\mathcal S}_F} = \left\{ {{\mathbf{F}}| {\mathop{\rm Tr}} \left[ {{{\mathbf{F}}^{\rm H}}{\mathbf{F}}} \right] \le {P_{\rm max}}} \right\},
	\end{equation}
	where ${P_{\rm{max}}}$ is the maximum transmit power of the BS.
	
	The IRS contains $M$ passive reflection elements that adjust the phases of incident signals.
	The set of reflection coefficients is represented as the vector $\bm{\phi}  = {\left[ {{\phi _1}, \cdots ,{\phi _M}} \right]^{\rm T}}$, or equivalently as a matrix of ${\bm{\Phi }} = diag\left( \bm{\phi}  \right)$, where ${\left| \phi _m \right|^2} = 1$, $\forall m = 1, \cdots ,M$.
	In order to provide efficient transmission, the antenna spacing at the BS should be large enough
	so that the small-scale fading associated with two different antennas can be assumed independent.
	A similar assumption holds for the reflection elements of the IRS.
	The baseband channels from the BS to the IRS, from the IRS to the BS, from user $k$ to the IRS, and from the IRS to user $k$ are denoted by ${\mathbf{G}_{\rm t}} \in {{\mathbb C}^{M \times {N_{\rm t}}}}$, ${\mathbf{G}_{\rm r}} \in {{\mathbb C}^{M \times {N_{\rm r}}}}$, ${\mathbf{h}_{{\rm t},k}} \in {{\mathbb C}^{M \times 1}}$, and ${\mathbf{h}_{{\rm r},k}} \in {{\mathbb C}^{M \times 1}}$, respectively.
	Furthermore, we denote the loop channels between the transmit and receive antenna(s) of user $k$ and the BS by $h_{kk}$ and ${{{\bf{H}}_{\rm B}}}$, respectively.
	The CSI for all channels is assumed to be quasi-static and perfectly known by the BS.
	\footnote{
		Though this assumption is idealistic, it allows us to explore the upper bounds for the performance of IRS-based FD networks.
		The robust transmission design based on imperfect CSI was studied in \cite{2020arXiv200801459W} for a multiuser half-duplex system, and its extension to FD systems will be left for future work.
	}

	The signal received by user $k$ can be modeled as
	\begin{align}\label{yDk}
		{y_{{\rm{D}},k}} &= {\bf{h}}_{{\rm r},k}^{\rm{H}}{\bf{\Phi }}{{\bf{G}}_{\rm t}}{{\bf{f}}_k}{s_{{\rm{D}},k}} + \underbrace {\sum\limits_{\substack{m = 1 \\ m \ne k}}^K {{\bf{h}}_{{\rm r},k}^{\rm{H}}{\bf{\Phi }}{{\bf{G}}_{\rm t}}{{\bf{f}}_m}{s_{{\rm{D}},m}}} }_{{\text{Multiuser interference}}} \notag \\
		&\quad + \underbrace {\sqrt {{\rho _{\rm{L}}}} \sqrt {{P_k}} {h_{kk}}{s_{{\rm{U}},k}}}_{{\text{Loop-interference}}}+ \underbrace {\sqrt {{\rho _{\rm{S}}}} \sqrt {{P_k}} {\bf{h}}_{{\rm r},k}^{\rm{H}}{\bf{\Phi }}{{\bf{h}}_{{\rm t},k}}{s_{{\rm{U}},k}}}_{{\text{Self-interference}}} \notag \\
		&\quad + \underbrace {\sum\limits_{\substack{m = 1 \\ m \ne k}}^K {\sqrt {{P_m}} {\bf{h}}_{{\rm r},k}^{\rm{H}}{\bf{\Phi }}{{\bf{h}}_{{\rm t},m}}{s_{{\rm{U}},m}}} }_{{\text{Co-channel interference}}} + {n_k},
	\end{align}
	where ${\rho}_{\rm L}$ and ${\rho_{\rm S}}$ with $0 \le {\rho_{\rm L}},{\rho_{\rm S}} \le 1$ are LI and SI coefficients, respectively,
	and $n_k$ is additive white Gaussian noise (AWGN) following the distribution ${\mathcal C}{\mathcal N}(0,\sigma _k^2)$.
	The coefficient $\rho_{\rm L}$ is introduced to model the fact that LI suppression methods such as antenna isolation may not completely eliminate the LI.
	Similarly, SI elimination methods can to some extent reduce the influence of SI reflected from the IRS,
	\footnote{
		According to \eqref{yDk}, to partially eliminate the SI, the scalar ${\bf{h}}_{{\rm r},k}^{\rm{H}}{\bf{\Phi }}{{\bf{h}}_{{\rm t},k}}$ should be estimated by each user,
		for example as follows.
		After the reflection coefficients of the IRS calculated at the BS are sent to the IRS controller, the BS remains silent and the IRS works with the calculated reflection coefficients.
		Then, each user sends one or more pilot symbols to estimate the scalar channel ${\bf{h}}_{{\rm r},k}^{\rm{H}}{\bf{\Phi }}{{\bf{h}}_{{\rm t},k}}$ while the other users remain silent.
		This step is repeated until all users have estimated their channels.
	}
	and thus we also introduce the coefficient $\rho_{\rm S}$ to model the residual SI component.
	Due to blockages as shown in Fig. \ref{Figsysmodel}, the user-to-user interference contribution will likely be small, and thus we treat it as AWGN and include it in $n_k$.
	In particular, we denote the sum of the LI term and $n_k$ in \eqref{yDk} as $i_{{\rm{D}},k}$, whose average power is given by $\sigma _{{\rm{D}},k}^2={\left| {{i_{{\rm{D}},k}}} \right|^2} = {{\rho}_{\rm{L}}} {P_k}{\left| {{h_{kk}}} \right|^2} + \sigma _k^2$.
	Then, the signal-to-interference-plus-noise ratio (SINR) at user $k$ is given by
	\begin{equation}
		{\gamma _{{\rm{D}},k}} = \frac{{{{\left| {{\bf{h}}_{{\rm r},k}^{\rm H}{\bf{\Phi }}{{\bf{G}}_{\rm t}}{{\bf{f}}_k}} \right|}^2}}}{{\sum\limits_{\substack{m = 1 \\ m \ne k}}^K {{{\left| {{\bf{h}}_{{\rm r},k}^{\rm H}{\bf{\Phi }}{{\bf{G}}_{\rm t}}{{\bf{f}}_m}} \right|}^2}}  +\sum\limits_{m = 1}^K {{\rho}}{{P_m}{{\left| {{\bf{h}}_{{\rm r},k}^{\rm H}{\bf{\Phi }}{{\bf{h}}_{{\rm t},m}}} \right|}^2}}  + \sigma _{{\rm{D}},k}^2}},
	\end{equation}
	where the coefficient ${\rho}$ is defined as
	\begin{equation}\nonumber
		{\rho}=\begin{cases}
			{\rho_{\rm{S}}}, & \text{if}\ m=k;\\
			1, & \text{otherwise}.
		\end{cases}
	\end{equation}
	
	\begin{figure*}[hb]
		\setcounter{equation}{12}
		\hrulefill
		\begin{align}\label{eUk1}
			{e_{{\rm{U}},k}}&=\mathbb E \left[ {{{\left( {{{\hat s}_{{\rm{U}},k}} - {s_{{\rm{U}},k}}} \right)}^{\rm{H}}}\left( {{{\hat s}_{{\rm{U}},k}} - {{\hat s}_{{\rm{U}},k}}} \right)} \right] \nonumber \\
			&= \left( {\sqrt {{P_k}} {\bf{u}}_{{\rm{U}},k}^{\rm H}{\bf{G}}_{\rm r}^{\rm H}{\bf{\Phi }}{{\bf{h}}_{{\rm t},k}} - 1} \right)^{\rm H}{\left( {\sqrt {{P_k}} {\bf{u}}_{{\rm{U}},k}^{\rm H}{\bf{G}}_{\rm r}^{\rm H}{\bf{\Phi }}{{\bf{h}}_{{\rm t},k}} - 1} \right)} + \sum\limits_{m = 1,m \ne k}^K {{P_m}{\bf{u}}_{{\rm{U}},k}^{\rm H}{\bf{G}}_{\rm r}^{\rm H}{\bf{\Phi }}{{\bf{h}}_{{\rm t},m}}{\bf{h}}_{{\rm t}.m}^{\rm H}{{\bf{\Phi }}^{\rm H}}{{\bf{G}}_{\rm r}}{{\bf{u}}_{{\rm{U}},k}}}  + \sigma _{\rm{U}}^2{N_{\rm r}}{\bf{u}}_{{\rm{U}},k}^{\rm H}{{\bf{u}}_{{\rm{U}},k}}\nonumber \\
			&= \sum\limits_{m = 1}^K {{P_m}{\bf{u}}_{{\rm{U}},k}^{\rm H}{\bf{G}}_{\rm r}^{\rm H}{\bf{\Phi }}{{\bf{h}}_{{\rm t},m}}{\bf{h}}_{{\rm t}.m}^{\rm H}{{\bf{\Phi }}^{\rm H}}{{\bf{G}}_{\rm r}}{{\bf{u}}_{{\rm{U}},k}}}  - 2{\mathop{\rm Re}\nolimits} \left\{ {\sqrt {{P_k}} {\bf{u}}_{{\rm{U}},k}^{\rm H}{\bf{G}}_{\rm r}^{\rm H}{\bf{\Phi }}{{\bf{h}}_{{\rm t},k}}} \right\} + \sigma _{\rm{U}}^2{\bf{u}}_{{\rm{U}},k}^{\rm H}{{\bf{u}}_{{\rm{U}},k}} + 1.
		\end{align}
	\end{figure*}

	\setcounter{equation}{5}
	Similarly, the signal received at the BS ${{\mathbf{y}}_{\rm U}} \in {{\mathbb C}^{{N_r} \times 1}}$ is given by
	\begin{align}\label{yU_origin}
		{{\bf{y}}_{\rm{U}}} &= {\bf{G}}_{\rm r}^{\rm H}{\bf{\Phi }}{{\bf{h}}_{{\rm t},k}}\sqrt {{P_k}} {s_{{\rm{U}},k}} + \underbrace {\sum\limits_{\substack{m = 1 \\ m \ne k}}^K {{\bf{G}}_{\rm r}^{\rm H}{\bf{\Phi }}{{\bf{h}}_{{\rm t},m}}\sqrt {{P_m}} {s_{{\rm{U}},m}}} }_{{\text{Multiuser interference}}} \notag\\
		&\quad + \underbrace { {{\bf{H}}_{\rm{B}}}\sum\limits_{m = 1}^K {{{\bf{f}}_m}} {s_{{\rm{D}},m}}}_{{\text{Loop-interference}}} + \underbrace {{\bf{G}}_{\rm r}^{\rm H}{\bf{\Phi }}{{\bf{G}}_{\rm t}}\sum\limits_{m = 1}^K {{{\bf{f}}_m}} {s_{{\rm{D}},m}}}_{{\text{Self-interference}}}+ {{\bf{n}}_{\rm{B}}},
	\end{align}
	where ${\bf{n}}_{\rm{B}}$ is the AWGN noise vector, whose elements are independently distributed as ${\cal C}{\cal N}(0,\sigma _{\rm{B}}^2)$.
	Based on techniques for LI cancellation for FD AF MIMO relays \cite{taghizadeh2018hardware,Riihonen2011Mitigation}, we assume the BS LI can be effectively eliminated.
	With the calculated reflection coefficients of the IRS, the SI received at the BS is known and can be effectively mitigated.
	We assume that any residual noise resulting from the interference cancellation is i.i.d. AWGN,
	denote $\sigma _{\rm{U}}^2$ as the average power of the total noise at the BS, and define ${i_n} \sim {\cal C}{\cal N}(0,\sigma _{\rm{U}}^2)$, $n=1,\dots,N_{\rm r}$.
	Then \eqref{yU_origin} can be simplified to
	\begin{align}\label{yU_simp}
		{{\bf{y}}_{\rm{U}}} = {\bf{G}}_{\rm r}^{\rm{H}}{\bf{\Phi }}{{\bf{h}}_{{\rm t},k}}\sqrt {{P_k}} {s_{{\rm{U}},k}} + \sum\limits_{\substack{m = 1 \\ m \ne k}}^K {{\bf{G}}_{\rm r}^{\rm{H}}{\bf{\Phi }}{{\bf{h}}_{{\rm t},m}}\sqrt {{P_m}} {s_{{\rm{U}},m}}}  + {{\bf{i}}_{\rm{B}}},
	\end{align}
	where ${{\bf{i}}_{\rm{B}}}  \triangleq  \left[ {i_1,\dots,i_{N_{\rm r}}} \right]^{\rm T}$.

	Denoting the set of receive beamformers at the BS by ${{\mathcal U}_{\rm{U}}} = {\left\{ {{{\bf{u}}_{{\rm{U}},k}}, \forall k \in {\mathcal K}} \right\}} $, the recovered signal for user $k$ is given by
	\begin{equation}\label{sUk}
		{\hat s_{{\rm{U}},k}} = {\bf{u}}_{{\rm{U}},k}^{\rm H}\left( {\sum\limits_{m = 1}^K {{\bf{G}}_{\rm r}^{\rm{H}}{\bf{\Phi }}{{\bf{h}}_{{\rm t},m}}\sqrt {{P_m}} {s_{{\rm{U}},m}}}  + {{\bf{i}}_{\rm{B}}}} \right).
	\end{equation}
	Then, the SINR of user $k$'s recovered signal is formulated as
	\begin{equation}
		{\gamma _{{\rm{U}},k}} = \frac{{{P_k}{{\left| {{\bf{u}}_{{\rm{U}},k}^{\rm H}{\bf{G}}_{\rm r}^{\rm H}{\bf{\Phi }}{{\bf{h}}_{{\rm t},k}}} \right|}^2}}}{{\sum\limits_{\substack{m = 1 \\ m \ne k}}^K {{P_m}{{\left| {{\bf{u}}_{{\rm{U}},k}^{\rm H}{\bf{G}}_{\rm r}^{\rm H}{\bf{\Phi }}{{\bf{h}}_{{\rm t},m}}} \right|}^2} + \sigma _{\rm{U}}^2{{\left| {{{\bf{u}}_{{\rm{U}},k}}} \right|}^2}} }}.
	\end{equation}
	{\spaceskip=0.19em\relax
		Accordingly, the maximum achievable rates (nat/s/Hz) of user $k$}
	{\spaceskip=0.25em\relax
		for downlink and uplink transmission are respectively given by}
	\begin{equation}\label{RDk}
		{R_{{\rm{D}},k}}\left( {{\bf{F}},{\bm{\phi}} } \right) = \log \left( {1 + {\gamma _{{\rm{D}},k}}} \right),
	\end{equation}
	and
	\begin{equation}\label{RUk}
		{R_{{\rm{U}},k}}\left( {\bm{\phi}} \right) = \log \left( {1 + {\gamma _{{\rm{U}},k}}} \right).
	\end{equation}
	
	\subsection{Problem Formulation}
	In this paper, we propose to guarantee the fairness among the users by maximizing the WMR by jointly optimizing the precoding matrix $\mathbf{F}$ and the reflection coefficient vector $\bm{\phi}$.
	Specifically, denoting $\omega _{l,k}\ge 1$ as a weighting factor, the WMR maximization problem is formulated as
	\begin{subequations}\label{P1}
		\begin{alignat}{2}
			\max_{{{\bf{F}},\bm{\phi} }} \quad & \min_{{l \in {\cal L},{k \in {\cal K}}}}{\left\{ {{\omega _{l,k}}{R_{l,k}}} \right\}} \\
			\mbox{s.t.}\quad
			&{\bf{F}} \in {{\cal S}_F},  \\
			&{\bm{\phi}}\in {{\cal S}_\phi },
		\end{alignat}
	\end{subequations}
	where the set ${{\cal S}_F}$ is defined in \eqref{pow_constraint}, and the set ${{\cal S}_{\bm \phi} } = \left\{ {\bm \phi | \left| {{\phi _m}} \right| = 1,1 \le m \le M} \right\}$ imposes the unit-modulus constraint on $\bm \phi$.
	\begin{remark}\label{remark:weight}
		Each weighting factor $\omega _{l,k}$ in the OF of Problem \eqref{P1} represents the inverse of the priority of the corresponding user.
		The optimal solution of Problem \eqref{P1} has a tendency to equalize the weighted rate of each user for both the uplink and downlink,
		which is consistent with our goal of ensuring fairness.
		However, the desired uplink and downlink rates in a cellular system are often asymmetric, so one may wish to choose weights that account for this difference.
		In particular, choosing a larger $\omega_{l,k}$ leads to a lower data rate for user $k$ in direction $l$.
	\end{remark}

	Note that Problem \eqref{P1} is difficult to solve as a result of the coupling between the precoding matrix $\bf F$ and the reflection coefficient vector $\bm{\phi}$, as well as the non-convex constraint on $\bm \phi$.
	In the following, efficient algorithms are provided to solve this problem.

	\section{SOCP-Based BCD Method} \label{SOCP-BASED BCD METHOD}
	In this section, we derive an efficient strategy for solving the formulated problem \eqref{P1}.
	We first rewrite \eqref{RDk} and \eqref{RUk} by using the equivalence between the WMR and the WMMSE to reformulate the original problem \eqref{P1} into a more tractable form \cite{9090356}, then optimize the subproblems relying on the block coordinate descent (BCD) algorithm framework.

	\subsection{Reformulation of the Original Problem}
	From \eqref{sUk}, the mean squared error (MSE) of the estimated signal at the BS corresponding to user $k$ can be derived as \eqref{eUk1} at the bottom of this page.
	\setcounter{equation}{13}
	Similarly, upon introducing the set of decoding variables ${{\mathcal U}_{\rm{D}}} = {\left\{ {u_{{\rm{D}},k}}, \forall k \in {\mathcal K} \right\}}$, the estimated signal symbol of user $k$ is given by ${\hat s_{{\rm{D}},k}} = u_{{\rm{D}},k}^ \ast {y_{{\rm{D}},k}}$.
	Then, the MSE of the estimated signal at user $k$ is written as \eqref{eDk1} at the bottom of the next page.
	\begin{figure*}[hb]
		\hrulefill
		\begin{align}\label{eDk1}
			{e_{{\rm{D}},k}} &=\mathbb E \left[ {{{\left( {{{\hat s}_{{\rm{D}},k}} - {s_{{\rm{D}},k}}} \right)}^{\rm H}}\left( {{{\hat s}_{{\rm{D}},k}} - {s_{{\rm{D}},k}}} \right)} \right]\notag \\
			&= {\left( {u_{{\rm{D}},k}^ * {\bf{h}}_{{\rm r},k}^{\rm H}{\bf{\Phi }}{{\bf{G}}_{\rm t}}{{\bf{f}}_k} - 1} \right)^{\rm H}}\left( {u_{{\rm{D}},k}^ * {\bf{h}}_{{\rm r},k}^{\rm H}{\bf{\Phi }}{{\bf{G}}_{\rm t}}{{\bf{f}}_k} - 1} \right) + \sum\limits_{m = 1,m \ne k}^K {u_{{\rm{D}},k}^ * {u_{{\rm{D}},k}}{\bf{h}}_{{\rm r},k}^{\rm H}{\bf{\Phi }}{{\bf{G}}_t}{{\bf{f}}_m}{\bf{f}}_m^{\rm H}{\bf{G}}_{\rm t}^{\rm H}{{\bf{\Phi }}^{\rm H}}{{\bf{h}}_{{\rm r},k}}}  \notag \\
			&\quad+\sum\limits_{m = 1}^K {{\rho}}{{P_m}u_{{\rm{D}},k}^ * {u_{{\rm{D}},k}}{\bf{h}}_{{\rm r},k}^{\rm H}{\bf{\Phi }}{{\bf{h}}_{{\rm r},m}}{\bf{h}}_{{\rm t},m}^{\rm H}{{\bf{\Phi }}^{\rm H}}{{\bf{h}}_{{\rm t},k}}}  + \sigma _{{\rm{D}},k}^2u_{{\rm{D}},k}^ * {u_{{\rm{D}}k}}\notag\\
			&= \sum\limits_{m = 1}^K {u_{{\rm{D}},k}^ * {u_{{\rm{D}},k}}{\bf{h}}_{{\rm r},k}^{\rm H}{\bf{\Phi }}{{\bf{G}}_{\rm t}}{{\bf{f}}_m}{\bf{f}}_m^{\rm H}{\bf{G}}_{\rm t}^{\rm H}{{\bf{\Phi }}^{\rm H}}{{\bf{h}}_{{\rm r},k}}}  - 2{\mathop{\rm Re}\nolimits} \left\{ {u_{{\rm{D}},k}^ * {\bf{h}}_{{\rm r},k}^{\rm H}{\bf{\Phi }}{{\bf{G}}_{\rm t}}{{\bf{f}}_k}} \right\} \notag\\
			&\quad +\sum\limits_{m = 1}^K {{\rho}}{{P_m}u_{{\rm{D}},k}^ * {u_{{\rm{D}},k}}{\bf{h}}_{{\rm r},k}^{\rm H}{\bf{\Phi }}{{\bf{h}}_{{\rm t},m}}{\bf{h}}_{{\rm t},m}^{\rm H}{{\bf{\Phi }}^{\rm H}}{{\bf{h}}_{{\rm r},k}}}  + \sigma _{{\rm{D}},k}^2u_{{\rm{D}},k}^ * {u_{{\rm{D}},k}} + 1.
		\end{align}
	\end{figure*}
	
	Introducing two sets of auxiliary variables: ${{\mathcal W}_{\rm{D}}} = \left\{ {{w_{{\rm{D}},k}} \ge 0,\forall k \in \mathcal K} \right\}$ and ${{\mathcal W}_{\rm{U}}} = \left\{ {{w_{{\rm{U}},k}} \ge 0,\forall k \in \mathcal K} \right\}$, the expressions for ${R_{{\rm{D}},k}}$ and ${R_{{\rm{U}},k}}$ can be transformed as follows
	\begin{align}
		{r_{{\rm{D}},k}}\left( {{\bf{F}},{\bm{\phi}},{{\mathcal U}_{\rm{D}}},{{\mathcal W}_{\rm{D}}} } \right) = \log \left| {{w_{{\rm{D}},k}}} \right| - {w_{{\rm{D}},k}}{e_{{\rm{D}},k}} + 1 \label{rDk},\\
		{r_{{\rm{U}},k}}\left( {{\bm {\phi}},{{\mathcal U}_{\rm{U}}},{{\mathcal W}_{\rm{U}}} } \right) = \log \left| {{w_{{\rm{U}},k}}} \right| - {w_{{\rm{U}},k}}{e_{{\rm{U}},k}} + 1 \label{rUk}.
	\end{align}
	Note that for a given reflection coefficient vector $\bm{\phi}$, ${r_{{\rm{D}},k}}\left( {{\bf{F}},{\bm{\phi}},{{\mathcal U}_{\rm{D}}},{{\mathcal W}_{\rm{D}}} } \right)$ and ${r_{{\rm{U}},k}}\left( {{{\bm {\phi}},{{\mathcal U}_{\rm{U}}},{{\mathcal W}_{\rm{U}}} }} \right)$ are concave functions for each set of variables when the others are fixed.
	Hence, we can reformulate Problem \eqref{P1} as
	\begin{subequations}\label{P2}
		\begin{alignat}{2}
			\max_{\substack{{{\mathcal U}_l}, {{\mathcal W}_l},{{l \in {\cal L}}}\\{\bf{F}},\bm{\phi} }} \quad & \min_{{{l \in {\cal L}}},{{k \in {\cal K}}}}{\left\{ {{\omega _{l,k}}{r_{l,k}}} \right\}} \\
			\mbox{s.t.}\quad
			&{\bf{F}} \in {{\cal S}_F} , \\
			&{\bm{\phi}}\in {{\cal S}_\phi }.
		\end{alignat}
	\end{subequations}
	
	Comparing the expressions of ${R_{{\rm{D}},k}}$ with ${r_{{\rm{D}},k}}$ and ${R_{{\rm{U}},k}}$ with ${r_{{\rm{U}},k}}$, the optimal ${\mathcal W}_{\rm{D}}$ and ${\mathcal W}_{\rm{U}}$ can be readily obtained as follows
	\begin{equation}\label{optw}
		{w_{{\rm{D}},k}} = e_{{\rm{D}},k}^{ - 1}, \quad {w_{{\rm{U}},k}} = e_{{\rm{U}},k}^{ - 1},\quad \forall k.
	\end{equation}
	For given ${\bf{F}}$, ${\bm{\phi}}$ and ${\mathcal W}_{\rm{D}}$, by setting the first-order derivative of ${r_{{\rm{D}},k}} \left( {{\bf{F}},{\bm{\phi}},{{\mathcal U}_{\rm{D}}},{{\mathcal W}_{\rm{D}}} } \right)$ with respect to (w.r.t.) ${u_{{\rm{D}},k}}$ to zero, we can obtain the optimal ${{\mathcal U}_{\rm{D}}}$ as shown in \eqref{optuD} at the bottom of the next page.
	\begin{figure*}[hb]
		\hrulefill
		\begin{equation}\label{optuD}
			{u_{{\rm{D}},k}} = {\bf{h}}_{{\rm r},k}^{\rm H}{\bf{\Phi }}{{\bf{G}}_{\rm t}}{{\bf{f}}_k}{\left( {\sum\limits_{m = 1}^K {{\bf{h}}_{{\rm r},k}^{\rm H}{\bf{\Phi }}{{\bf{G}}_{\rm t}}{{\bf{f}}_m}{\bf{f}}_m^{\rm H}{\bf{G}}_{\rm t}^{\rm H}{{\bf{\Phi }}^{\rm H}}{{\bf{h}}_{{\rm r},k}}}  +\sum\limits_{m = 1}^K {{\rho}}{{P_m}{\bf{h}}_{{\rm r},k}^{\rm H}{\bf{\Phi }}{{\bf{h}}_{{\rm t},m}}{\bf{h}}_{{\rm t},m}^{\rm H}{{\bf{\Phi }}^{\rm H}}{{\bf{h}}_{{\rm r},k}}}  + \sigma _{{\rm{D}},k}^2} \right)^{ - 1}}.
		\end{equation}
	\end{figure*}
	Similarly, the optimal linear receivers in ${\mathcal U}_{\rm{U}}$ can be derived by setting the first-order derivative of ${r_{{\rm{U}},k}}\left( {{\bm {\phi}},{{\mathcal U}_{\rm{U}}},{{\mathcal W}_{\rm{U}}} } \right)$ w.r.t ${u_{{\rm{U}},k}}$ to zero, as follows
	\vspace{-0.2 cm}
		\begin{align}\label{optuU}
			{{\bf{u}}_{{\rm{U}},k}} &= \sqrt {{P_k}} \left({{\sum\limits_{m = 1}^K {{P_m}{\bf{G}}_{\rm{r}}^{\rm{H}}{\bf{\Phi }}{{\bf{h}}_{{\rm{t}},m}}{\bf{h}}_{{\rm{t}},m}^{\rm{H}}{{\bf{\Phi }}^{\rm{H}}}{{\bf{G}}_{\rm{r}}}}  + \sigma _{\rm{U}}^2{{\bf{I}}_{{N_{\rm{r}}}}}}}\right)^{-1}\notag\\
			&\quad\cdot{ {\bf{G}}_{\rm{r}}^{\rm{H}}{\bf{\Phi }}{{\bf{h}}_{{\rm{t}},k}}}.
		\end{align}
	\vspace{-0.6 cm}
	
	In the following, we adopt the BCD method to solve Problem \eqref{P2} by alternately optimizing the OF over each of the variables.
	Since the optimal ${\mathcal U}_{\rm{D}}$, ${\mathcal W}_{\rm{D}}$, ${\mathcal U}_{\rm{U}}$ and ${\mathcal W}_{\rm{U}}$ in each iteration are given by \eqref{optw}-\eqref{optuU}, the main task is the optimization of the precoding matrix $\bf F$ and the reflection coefficient vector $\bm{\phi}$.
	
	\subsection{Optimizing the Precoding Matrix $\bf F$}
	Note that the precoding matrix $\bf F$ is not related to the rate of the uplink transmission $r_{{\rm{U}},k}$, so to optimize $\bf F$ for a given $\bm{\phi}$, we can simplify the OF of Problem \eqref{P2} to
	\begin{equation}
		\min{\left\{ {{\omega _{{\rm{D}},k}}{r_{{\rm{D}},k}\left( {{\bf{F}}} \right)}} \right\}}.
	\end{equation}
	
	We introduce a selection vector ${\bf t}_k \in {\mathbb R}^{K \times 1}$, in which all elements are zero except the $k$th one.
	Then, from \eqref{eDk1}, we have
	\begin{align}\label{eDk2}
		{e_{{\rm{D}},k}} &=\sum\limits_{m = 1}^K {u_{{\rm{D}},k}^ * {u_{{\rm{D}},k}}{{\left( {{\bf{F}}{{\bf{t}}_m}} \right)}^{\rm H}}{\bf{G}}_{\rm t}^{\rm H}{{\bf{\Phi }}^{\rm H}}{{\bf{h}}_{{\rm r},k}}{\bf{h}}_{{\rm r},k}^{\rm H}{\bf{\Phi }}{{\bf{G}}_{\rm t}}{\bf{F}}{{\bf{t}}_m}}  \nonumber \\
		&\quad - 2{\mathop{\rm Re}\nolimits} \left\{ {u_{{\rm{D}},k}^ * {\bf{h}}_{{\rm r},k}^{\rm H}{\bf{\Phi }}{{\bf{G}}_{\rm t}}{\bf{F}}{{\bf{t}}_k}} \right\} \nonumber \\
		&\quad +\sum\limits_{m = 1}^K {{\rho}}{{P_m}u_{{\rm{D}},k}^ * {u_{{\rm{D}},k}}{\bf{h}}_{{\rm r},k}^{\rm H}{\bf{\Phi }}{{\bf{h}}_{{\rm t},m}}{\bf{h}}_{{\rm t},m}^{\rm H}{{\bf{\Phi }}^{\rm H}}{{\bf{h}}_{{\rm r},k}}}  \nonumber \\
		&\quad + \sigma _{{\rm{D}},k}^2u_{{\rm{D}},k}^ * {u_{{\rm{D}},k}} + 1\nonumber \\
		& = {\mathop{\rm Tr}\nolimits} \left[ {u_{{\rm{D}},k}^ * {u_{{\rm{D}},k}}{{\bf{F}}^{\rm H}}{\bf{G}}_{\rm t}^{\rm H}{{\bf{\Phi }}^{\rm H}}{{\bf{h}}_{{\rm r},k}}{\bf{h}}_{{\rm r},k}^{\rm H}{\bf{\Phi }}{{\bf{G}}_{\rm t}}{\bf{F}}} \right] \nonumber \\
		&\quad - 2{\mathop{\rm Re}\nolimits} \left\{ {{\mathop{\rm Tr}\nolimits} \left[ {u_{{\rm{D}},k}^ * {\bf{h}}_{{\rm r},k}^{\rm H}{\bf{\Phi }}{{\bf{G}}_{\rm t}}{\bf{F}}{{\bf{t}}_k}} \right]} \right\} \nonumber \\
		&\quad +\sum\limits_{m = 1}^K {{\rho}}{{P_m}u_{{\rm{D}},k}^ * {u_{{\rm{D}},k}}{\bf{h}}_{{\rm r},k}^{\rm H}{\bf{\Phi }}{{\bf{h}}_{{\rm t},m}}{\bf{h}}_{{\rm t},m}^{\rm H}{{\bf{\Phi }}^{\rm H}}{{\bf{h}}_{{\rm r},k}}}  \nonumber \\
		&\quad + \sigma _{{\rm{D}},k}^2u_{{\rm{D}},k}^ * {u_{{\rm{D}},k}} + 1.
	\end{align}
	Substituting \eqref{eDk2} into \eqref{rDk} and defining ${h_{{\rm{D}},k}}\left( {\bf{F}} \right)  =  {\omega _{{\rm{D}},k}}{r_{{\rm{D}},k}}\left( {\bf{F}} \right)$, $\forall k \in \mathcal K$, we formulate the subproblem for the optimization of $\bf{F}$ from Problem \eqref{P2} as
	\begin{subequations}\label{PF1}
		\begin{alignat}{2}
			\max_{\bf{F}} \quad & \min_{{{k \in {\cal K}}}}{\left\{ {h_{{\rm{D}},k}}\left( {\bf{F}} \right) \right\}} \\
			\mbox{s.t.}\quad
			&{\bf{F}} \in {{\cal S}_F}.
		\end{alignat}
	\end{subequations}
	
	It can be derived that
	\begin{equation}
		{h_{{\rm{D}},k}}\left( {\bf{F}} \right)=2{\mathop{\rm Re}\nolimits} \left\{ {{\mathop{\rm Tr}\nolimits} \left[ {{\bf{C}}_k^{\rm H}{\bf{F}}} \right]} \right\} - {\mathop{\rm Tr}\nolimits} \left[ {{{\bf{F}}^{\rm H}}{{\bf{B}}_k}{\bf{F}}} \right] + {{\rm{const}}_k},
	\end{equation}
	where ${{\bf{B}}_k}$, ${{\bf{C}}_k}$ and ${{\rm{const}}_k}$ are respectively given by
	\begin{equation}\nonumber
		\begin{aligned}
			&{{\bf{B}}_k}\triangleq {\omega _{{\rm{D}},k}}{w_{{\rm{D}},k}}u_{{\rm{D}},k}^ * {u_{{\rm{D}},k}}{\bf{G}}_{\rm t}^{\rm H}{{\bf{\Phi }}^{\rm H}}{{\bf{h}}_{{\rm r},k}}{\bf{h}}_{{\rm r},k}^{\rm H}{\bf{\Phi }}{{\bf{G}}_{\rm t}}, \\
			&{{\bf{C}}_k}\triangleq \omega _{{\rm{D}},k}^ * w_{{\rm{D}},k}^ * {u_{{\rm{D}},k}}{\bf{G}}_{\rm t}^{\rm H}{{\bf{\Phi }}^{\rm H}}{{\bf{h}}_{{\rm r},k}}{\bf{t}}_k^{\rm H},\\
			&{\rm{cons}}{{\rm{t}}_k}\\
			&\triangleq {\omega _{{\rm{D}},k}}\log \left| {{w_{{\rm{D}},k}}} \right| + {\omega _{{\rm{D}},k}} + {\omega _{{\rm{D}},k}}{w_{{\rm{D}},k}}\left( {\sigma _{{\rm{D}},k}^2u_{{\rm{D}},k}^*{u_{{\rm{D}},k}} + 1} \right) \\
			&\quad - {\omega _{{\rm{D}},k}}{w_{{\rm{D}},k}}\sum\limits_{m = 1}^K \rho  {P_m}u_{{\rm{D}},k}^*{u_{{\rm{D}},k}}{\bf{h}}_{{\rm{r}},k}^{\rm{H}}{\bf{\Phi }}{{\bf{h}}_{{\rm{t}},m}}{\bf{h}}_{{\rm{t}},m}^{\rm{H}}{{\bf{\Phi }}^{\rm{H}}}{{\bf{h}}_{{\rm{r}},k}}.
		\end{aligned}
	\end{equation}
	Then, by introducing auxiliary variable $\delta$ for the pointwise minimum expressions, Problem \eqref{PF1} can be reformulated as follows
	\begin{subequations}\label{PF2}
		\begin{alignat}{2}
			\max_{{\bf{F}},\delta} \quad & \delta \\
			\mbox{s.t.}\quad
			&{h_{{\rm{D}},k}}\left( {\bf{F}} \right) \ge \delta ,\forall k \in \mathcal K,\label{PF2_rate_constraint}\\
			&{\bf{F}} \in {{\cal S}_F}.\label{PF2_pow_constraint}
		\end{alignat}
	\end{subequations}
	Problem \eqref{PF2} is an SOCP, which can be optimally solved by the existing optimization tools, such as CVX.
	
	\subsection{Optimizing the Reflection Coefficient Vector $\bm \phi$}
	In this subsection, we optimize $\bm \phi$ given $\bf{F}$.
	Defining
	\[{{\bf{\tilde H}}_{{\rm r},k}} \triangleq u_{{\rm{D}},k}^ * {u_{{\rm{D}},k}}{{\bf{h}}_{{\rm r},k}}{\bf{h}}_{{\rm r},k}^{\rm H},\]
	\[{{\bf{\tilde G}}_{\rm t}} \triangleq \sum\limits_{m = 1}^K {{{\bf{G}}_{\rm t}}{{\bf{f}}_m}{\bf{f}}_m^{\rm H}{\bf{G}}_{\rm t}^{\rm H}},\]
	\[{{\bf{\tilde H}}_{{\rm t},k}} \triangleq \sum\limits_{m = 1}^K {{\rho}{P_m}{{\bf{h}}_{{\rm t},m}}{\bf{h}}_{{\rm t},m}^{\rm H}},\]
	we can reformulate \eqref{eDk1} as
	\begin{align}\label{eDk4}
		{e_{{\rm{D}},k}} &= {\mathop{\rm Tr}\nolimits} \left[ {{{\bf{\Phi }}^{\rm H}}{{{\bf{\tilde H}}}_{{\rm r},k}}{\bf{\Phi }}{{{\bf{\tilde G}}}_{\rm t}} + {{\bf{\Phi }}^{\rm H}}{{{\bf{\tilde H}}}_{{\rm r},k}}{\bf{\Phi }}{{{\bf{\tilde H}}}_{{\rm t},k}}} \right] \nonumber\\
		&\quad - 2{\mathop{\rm Re}\nolimits} \left\{ {{\mathop{\rm Tr}\nolimits} \left[ {u_{{\rm{D}},k}^ * {{\bf{G}}_{\rm t}}{{\bf{f}}_k}{\bf{h}}_{{\rm r},k}^{\rm H}{\bf{\Phi }}} \right]} \right\} + \sigma _{{\rm{D}},k}^2u_{{\rm{D}},k}^ * {u_{{\rm{D}},k}} + 1 \nonumber\\
		&= {{\bm{\phi}} ^{\rm H}}\left( {{{{\bf{\tilde H}}}_{{\rm r},k}} \odot {{\left( {{{{\bf{\tilde G}}}_{\rm t}} + {{{\bf{\tilde H}}}_{{\rm t},k}}} \right)}^{\rm T}}} \right){\bm{\phi}}  \nonumber \\
		&\quad - 2{\mathop{\rm Re}\nolimits} \left\{ {{\bf{g}}_{{\rm{D}},k}^{\rm T}{\bm{\phi}} } \right\} + \sigma _{{\rm{D}},k}^2u_{{\rm{D}},k}^ * {u_{{\rm{D}},k}} + 1,
	\end{align}
	where ${{\bf{g}}_{{\rm{D}},k}}$ is the collection of diagonal elements of the matrix ${\left[ {u_{{\rm{D}},k}^ * {{\bf{G}}_{\rm t}}{{\bf{f}}_k}{\bf{h}}_{{\rm r},k}^{\rm H}} \right]}$ \cite[Eq. (1.10.6)]{Zhang2017Matrix}, i.e.
	\[{{\bf{g}}_{{\rm{D}},k}} \triangleq {\left[ {{{\left[ {u_{{\rm{D}},k}^ * {{\bf{G}}_{\rm t}}{{\bf{f}}_k}{\bf{h}}_{{\rm r},k}^{\rm H}} \right]}_{1,1}}, \dots ,{{\left[ {u_{{\rm{D}},k}^ * {{\bf{G}}_{\rm t}}{{\bf{f}}_k}{\bf{h}}_{{\rm r},k}^{\rm H}} \right]}_{M,M}}} \right]^{\rm T}}.\]
	
	Similarly, from \eqref{eUk1}, we have
	\begin{align}\label{eUk2}
		{e_{{\rm{U}},k}} &= {\mathop{\rm Tr}\nolimits} \left[ {{{\bf{\Phi }}^{\rm H}}{{{\bf{\tilde G}}}_{{\rm r},k}}{\bf{\Phi }}{{{\bf{\tilde H}}}_{{\rm t}}}} \right]- 2{\mathop{\rm Re}\nolimits} \left\{ {{\mathop{\rm Tr}\nolimits} \left[{\sqrt {{P_k}} {{\bf{h}}_{{\rm t},k}}{\bf{u}}_{{\rm{U}},k}^{\rm H}{\bf{G}}_{\rm r}^{\rm H}{\bf{\Phi }}} \right]} \right\}\nonumber\\
		&\quad + \sigma _{\rm{U}}^2{\bf{u}}_{{\rm{U}},k}^{\rm H}{{\bf{u}}_{{\rm{U}},k}} + 1\nonumber\\
		&= {{\bm{\phi}} ^{\rm H}}\left( {{{{\bf{\tilde G}}}_{{\rm r},k}} \odot {\bf{\tilde H}}_{{\rm t}}^{\rm T}} \right){\bm{\phi}}  - 2{\mathop{\rm Re}\nolimits} \left\{ {{\bf{g}}_{{\rm{U}},k}^{\rm T}{\bm{\phi}} } \right\} \nonumber\\
		&\quad + \sigma _{\rm{U}}^2{\bf{u}}_{{\rm{U}},k}^{\rm H}{{\bf{u}}_{{\rm{U}},k}} + 1,
	\end{align}
	where
	\[{{\bf{\tilde G}}_{{\rm r},k}} \triangleq{{\bf{G}}_{\rm r}}{{\bf{u}}_{{\rm{U}},k}}{\bf{u}}_{{\rm{U}},k}^{\rm H}{\bf{G}}_{\rm r}^{\rm H},\]
	\[{{\bf{\tilde H}}_{{\rm t}}} \triangleq \sum\limits_{m = 1}^K {{P_m}{{\bf{h}}_{{\rm t},m}}{\bf{h}}_{{\rm t},m}^{\rm H}},\]
	and vector ${\bf{g}}_{{\rm{U}},k}$ is the collection of diagonal elements of the matrix $\left[ {\sqrt {{P_k}} {{\bf{h}}_{{\rm t},k}}{\bf{u}}_{{\rm{U}},k}^{\rm H}{\bf{G}}_{\rm r}^{\rm H}} \right]$.
	
	Define ${h_{l,k}}\left( {\bm{\phi}} \right) = {\omega _{l,k}}{r_{l,k}}\left( {\bm{\phi}} \right)$ for $\forall l \in \mathcal L, k \in \mathcal K$.
	Substituting \eqref{eDk4} and \eqref{eUk2} into \eqref{rDk} and \eqref{rUk}, respectively, it can be derived that
	\begin{equation}
		{h_{l,k}}\left( {\bm{\phi}}  \right) = 2{\mathop{\rm Re}\nolimits} \left\{ {{\bf{a}}_{l,k}^{\rm H}{\bm{\phi}} } \right\} - {{\bm{\phi}} ^{\rm H}}{{\bf{A}}_{l,k}}{\bm{\phi}}  + {{\rm{const}}_{l,k}},
	\end{equation}
	where ${{\bf{a}}_{l,k}}$, ${{\bf{A}}_{l,k}}$ and ${{\rm{const}}_{l,k}}$ are respectively given by
	\begin{equation}\nonumber
		\begin{aligned}
			&{{\bf{a}}_{l,k}}\triangleq \omega _{l,k}^ * w_{l,k}^ * {\bf{g}}_{l,k}^ *, \; \forall l \in \mathcal L,\\
			&{{\bf{A}}_{{\rm{D}},k}}\triangleq {\omega _{{\rm{D}},k}}{w_{{\rm{D}},k}}{{\bf{\tilde H}}_{{\rm r},k}} \odot {\left( {{{{\bf{\tilde G}}}_{\rm t}} + {{{\bf{\tilde H}}}_{{\rm t},k}}} \right)^{\rm T}},\\
			&{{\bf{A}}_{{\rm{U}},k}}\triangleq {\omega _{{\rm{U}},k}}{w_{{\rm{U}},k}}{{\bf{\tilde G}}_{{\rm r},k}} \odot {\bf{\tilde H}}_{{\rm t}}^{\rm T},\\
			&{{\rm{const}}_{{\rm{D}},k}}\triangleq {\omega _{{\rm{D}},k}}\left( {\log \left| {{w_{{\rm{D}},k}}} \right| + 1} \right) \\
			&\qquad\qquad\quad - {\omega _{{\rm{D}},k}}{w_{{\rm{D}},k}}\left( {\sigma _{{\rm{D}},k}^2u_{{\rm{D}},k}^*{u_{{\rm{D}},k}} + 1} \right),\\
			&{{\rm{const}}_{{\rm{U}},k}}\triangleq {\omega _{{\rm{U}},k}}\left( {\log \left| {{w_{{\rm{U}},k}}} \right| + 1} \right)\\
			&\qquad\qquad\quad - {\omega _{{\rm{U}},k}}{w_{{\rm{U}},k}}\left( {\sigma _{\rm{U}}^2{\bf{u}}_{{\rm{U}},k}^{\rm{H}}{{\bf{u}}_{{\rm{U}},k}} + 1} \right).
		\end{aligned}
	\end{equation}
	Then, the subproblem for the optimization of $\bm \phi$ is formulated as
	\vspace{-0.2cm}
	\begin{subequations}\label{Pphi1}
		\begin{alignat}{2}
			\max_{\bm{\phi}} \quad & \min_{{{l \in {\cal L}},{k \in {\cal K}}}}{\left\{ {h_{l,k}}\left( {\bm{\phi}} \right) \right\}} \\
			\mbox{s.t.}\quad
			&{\bm{\phi}} \in {{\cal S}_{\phi}}.
		\end{alignat}
	\end{subequations}
	Introducing auxiliary variable $\epsilon$, Problem \eqref{Pphi1} is equivalent to
	\vspace{-0.3cm}
	\begin{subequations}\label{Pphi2}
		\begin{alignat}{2}
			\max_{{\bm{\phi}},\epsilon} \quad & \epsilon \\
			\mbox{s.t.}\quad
			&{h_{l,k}}\left( {\bm{\phi}} \right) \ge \epsilon ,\forall l \in \mathcal L,\forall k \in \mathcal K,\\
			&{\bm{\phi}} \in {{\cal S}_{\phi}}.\label{Pphi2c}
		\end{alignat}
	\end{subequations}
	
	Problem \eqref{Pphi2} is still non-convex, due to the unit-modulus constraint \eqref{Pphi2c}.
	To address this issue, we take a straightforward approach that replacing ${\cal S}_{{ \phi}}$ with the relaxed constraint set ${{\cal S}_{{ \phi}}^{\rm {relax}}} = \left\{ {\bm \phi | \left| {{\phi _m}} \right| \le 1,1 \le m \le M} \right\}$,
	then Problem \eqref{Pphi2} is transformed into an SOCP which can be optimally solved.
	Denote the optimal solution of the relaxed version of Problem \eqref{Pphi2} by $\tilde {\bm{\phi}}$.
	Then, a proximate optimal solution for the original Problem \eqref{Pphi2} can be obtained by $\hat {\bm{\phi}}=\exp \left\{ {j\angle {\tilde {\bm{\phi}}}} \right\}$, where $\angle \left(  \cdot  \right)$ and $\exp \left\{  \cdot  \right\}$ are both element-wise operations.
	Note that the global optimality of $\hat {\bm{\phi}}$ obtained may not be guaranteed at each iteration.
	To ensure the convergence, we adopt the following strategy:
	\begin{equation}\label{phi_strtegy}
		{\bm{\phi}}=\begin{cases}
			\hat {\bm{\phi}}, & \text{if}\ \min\limits_{{{l },{k }}}{\left\{ {h_{l,k}}\left( {{\hat{\bm{\phi}}}} \right) \right\}} \ge \min\limits_{{{l},{k }}}{\left\{ {h_{l,k}}\left( {{\bm{\phi}}} \right) \right\}};\\
			{\bm{\phi}}, & \text{otherwise}.
		\end{cases}
	\end{equation}

	\subsection{Algorithm Development}
	\subsubsection{SOCP based BCD algorithm}
	Based on the discussions above, we provide the details of the proposed BCD algorithm in Algorithm \ref{SCOP-based BCD},
	where the optimization variables ${\mathcal U}_{\rm D}$, ${\mathcal U}_{\rm U}$, ${\mathcal W}_{\rm D}$, ${\mathcal W}_{\rm U}$, $\bf F$ and $\bm \phi$ are alternately updated to maximize the WMR of all users.

	%
	In Algorithm \ref{SCOP-based BCD}, the globally optimal solution to Problem \eqref{PF2} can be obtained at each iteration.
	While the adopted relaxation technique leads to some performance loss in solving Problem \eqref{Pphi2},
	which mainly depends on the approximation gap between $\hat {\bm{\phi}}$ and ${\tilde {\bm{\phi}}}$.
	Hence, the optimality of Algorithm \ref{SCOP-based BCD} is not guaranteed.
	However, the simulation results in Section \ref{SIMULATION RESULTS} illustrates that the performance loss is actually limited when the IRS is deployed at the BS side.

	\renewcommand{\algorithmicrequire}{\textbf{Initialize:}}
	\renewcommand{\algorithmicensure}{\textbf{Output:}}
	
	\begin{algorithm}[t] 
		\caption{SOCP based BCD algorithm} 
		\label{SCOP-based BCD} 
		\begin{algorithmic}[1] 
			\Require  \mbox{Initial iteration number $n=1$, and feasible ${\bf F}^1$, ${\bm \phi}^1$.}
			\Repeat
			\State Given ${\bf F}^{n}$ and ${\bm \phi}^n$, calculate the optimal decoding variables ${\mathcal U}_{\rm{D}}^{n+1}$ in \eqref{optuD} and the optimal linear receivers ${\mathcal U}_{\rm{U}}^{n+1}$ in \eqref{optuU};
			\State Given ${\bf F}^{n}$, ${\bm \phi}^n$, ${\mathcal U}_{\rm{D}}^{n+1}$ and ${\mathcal U}_{\rm{U}}^{n+1}$, calculate the optimal auxiliary variables ${\mathcal W}_{\rm{D}}^{n+1}$ and ${\mathcal W}_{\rm{U}}^{n+1}$ in \eqref{optw};
			\State Given ${\mathcal U}_{\rm{D}}^{n+1}$, ${\mathcal U}_{\rm{U}}^{n+1}$, ${\mathcal W}_{\rm{D}}^{n+1}$, ${\mathcal W}_{\rm{U}}^{n+1}$ and ${\bm \phi}^n$, calculate the optimal precoding matrix ${\bf F}^{n+1}$ by solving \mbox{Problem \eqref{PF2};}\label{update_F_BCD}
			\State Given ${\mathcal U}_{\rm{D}}^{n+1}$, ${\mathcal U}_{\rm{U}}^{n+1}$, ${\mathcal W}_{\rm{D}}^{n+1}$, ${\mathcal W}_{\rm{U}}^{n+1}$ and ${\bf F}^{n+1}$, calculate the optimal reflection coefficient vector ${\bm \phi}^{n+1}$ by solving Problem \eqref{Pphi2};\label{update_phi_BCD}
			\State Set $n \leftarrow n+1$;
			\Until{The value of the OF in \eqref{P2} converges.}
		\end{algorithmic}
	\end{algorithm}
	
	\subsubsection{Complexity Analysis}\label{comp_analy_BCD}
	First, we have to compute the value of ${\mathcal U}_D$, ${\mathcal U}_U$, ${\mathcal W}_D$, and ${\mathcal W}_U$.
	The computational complexity of this step is analysed as follows:
	The order of complexity for computing each $u_{{\rm D},k}$ in \eqref{optuD} and each $u_{{\rm U},k}$ in \eqref{optuU} is given by ${\cal O}\left(K\left(M^2+N_{\rm t} M\right)\right)$ and ${\cal O}\left(K\left(M^2+N_{\rm r} M\right)+M^3\right)$, respectively.
	The complexity order of computing ${\mathcal U}_D$ and ${\mathcal U}_U$ is ${\cal O}\left(K^2\left(M^2+N_{\rm t} M+N_{\rm r} M\right)+KM^3\right)$.
	The complexity of computing ${\mathcal W}_D$ and ${\mathcal W}_U$ is equal to that of computing the $K$ values of $e_{{\rm D},k}$ in \eqref{eDk1} of order ${\cal O}\left(K\left(M^2+N_{\rm t} M\right)\right)$ and the $K$ values of $e_{{\rm U},k}$ in \eqref{eUk1} of order ${\cal O}\left(K\left(M^2+N_{\rm r} M\right)\right)$, respectively.
	Thus, the overall complexity of computing ${\mathcal W}_{\rm{D}}$ and ${\mathcal W}_{\rm{U}}$ is ${\cal O}\left(K^2\left(M^2+N_{\rm t} M+N_{\rm r} M\right)\right)$,
	and the total complexity is of order ${\cal O}\left(K^2\left(M^2+N_{\rm t} M+N_{\rm r} M\right)+KM^3\right)$.
	
	Then, we analyse the complexity of solving the two SOCPs in steps \ref{update_F_BCD} and \ref{update_phi_BCD}.
	Problem \eqref{PF2} contains $K$ rate constraints in \eqref{PF2_rate_constraint} and a power constraint in \eqref{PF2_pow_constraint}.
	Since each of the constraints is of dimension $KN_{\rm t}$, the total complexity is of order ${\cal O}\left(K^{5.5} N^3_{\rm t}\right)$ \cite{ben2001lectures}.
	Similarly, the complexity of solving the relaxed version of Problem  \eqref{Pphi2}, which contains $2K$ rate constraints with dimension $M$ and $M$ constant modulus constraint with dimension one, is of order ${\cal O}\left(M^{3.5} + M\left(2K\right)^{3.5}+M^3\left(2K\right)^{2.5}\right)$.
	As a result, the total order of the complexity for Algorithm \ref{SCOP-based BCD} per iteration is given by
	\begin{equation}\label{complex_Alg1}
		{\cal C}_{\rm {Alg. \ref{SCOP-based BCD}}}={\cal O}\left(M^{3.5}+ M^3 K^{2.5} + K^{5.5} N^3_{\rm t}\right),
	\end{equation}
	which is dominated by the complexity of solving Problem \eqref{PF2} and \eqref{Pphi2}.

	\section{Low-Complexity Algorithm Development} \label{LOW-COMPLEXITY ALGORITHM DEVELOPMENT}
	In Algorithm 1, there are an SOCP and a quasi-SOCP that have to be solved in each BCD iteration.
	To reduce the computational load, in this section we propose a low-complexity algorithm with closed-form solutions.
	Since the OFs of Problem \eqref{PF1} and \eqref{Pphi1} are non-differentiable, we first derive a lower-bound approximation by introducing a smooth approximation \cite{Xu2001Smoothing}.
	The approximated problem is then solved using the MM method.
	
	The following two smoothing functions $f\left( {\bf{F}} \right)$ and $f\left( {\bm{\phi}}  \right)$ are introduced to approximate the OFs of Problem \eqref{PF1} and \eqref{Pphi1}, respectively:
	\begin{align}\label{f(F)}
		&\min_{{{k \in {\cal K}}}}{\left\{ {h_{{\rm{D}},k}}\left( {\bf{F}} \right) \right\}}\notag\\
		&\qquad \approx f\left( {\bf{F}} \right)=  - \frac{1}{\mu }\log \left( {\sum\limits_{k \in {\cal K}} {\exp \left\{ { - \mu {h_{{\rm{D}},k}}\left( {\bf{F}} \right)} \right\}} } \right),
	\end{align}
	\begin{align}\label{f(phi)}
		&\min_{{{l \in {\cal L}},{k \in {\cal K}}}}{\left\{ {h_{l,k}}\left( {\bm{\phi}} \right) \right\}} \notag \\
		&\quad \approx f\left( {\bm{\phi}}  \right)=  - \frac{1}{\mu }\log \left( {\sum\limits_{l \in {\cal L}} {\sum\limits_{k \in {\cal K}} {\exp \left\{ { - \mu {h_{l,k}}\left( {\bm{\phi}}  \right)} \right\}} } } \right),
	\end{align}
	where $\mu > 0$ is a smoothing parameter.
	For $\mu > 0$, the following inequalities hold:
	\begin{equation}\label{smoothchar_F}
		f\left( {\bf{F}} \right) \le \min_{{{k \in {\cal K}}}}{\left\{ {h_{{\rm{D}},k}}\left( {\bf{F}} \right) \right\}} \le f\left( {\bf{F}} \right) + \frac{1}{\mu }\log \left( K \right)
	\end{equation}
	\begin{equation}\label{smoothchar_phi}
		f\left( {\bm{\phi}}  \right) \le \min_{{{l \in {\cal L}},{k \in {\cal K}}}}{\left\{ {h_{l,k}}\left( {\bm{\phi}} \right) \right\}} \le f\left( {\bm{\phi}}  \right) + \frac{1}{\mu }\log \left( {2K} \right).
	\end{equation}
	
	As shown in \eqref{smoothchar_F} and \eqref{smoothchar_phi}, $f\left( {\bf{F}} \right)$ and $f\left( {\bm{\phi}}  \right)$ are the lower-bounds for the OFs of Problem \eqref{PF1} and \eqref{Pphi1}, respectively.
	Moreover, it has been proved in \cite{zhou2019intelligent} that function $-\frac{1}{\mu}\log \left( \sum_{k\in \mathcal{K}}{\exp \left\{ -\mu x_k \right\}} \right) $ is increasing and concave w.r.t. $x_k$.
	Note that quadratic functions ${h_{{\rm{D}},k}}\left( {\bf{F}} \right)$ and ${h_{l,k}}\left( {\bm{\phi}}  \right)$ are concave w.r.t. ${\bf{F}}$ and ${\bm{\phi}}$, respectively,
	so $f \left( {\bf{F}} \right)$ and $f\left( {\bm{\phi}}  \right)$ are concave functions w.r.t. ${\bf{F}}$ and ${\bm{\phi}}$, respectively.
	
	Recall that $\min_{{{k \in {\cal K}}}}{\left\{ {h_{{\rm{D}},k}}\left( {\bf{F}} \right) \right\}}$ and $\min_{{{l \in {\cal L}},{k \in {\cal K}}}}{\left\{ {h_{l,k}}\left( {\bm{\phi}} \right) \right\}}$ are piecewise functions and non-differentiable, which is the reason why we adopt the smoothing method.
	Thus, the strategy of initializing and adjusting $\mu$ should be chosen appropriately.
	On the one hand, in the early stage of the BCD algorithm, a large $\mu$ may trap ${\bf F}^n$ and ${\bm \phi}^n$ in a local stationary point far from the optimal solutions of Problem \eqref{PF1} and \eqref{Pphi1}.
	On the other hand, in order to make the algorithm converge to globally optimal solutions, a large $\mu$ is required to improve the approximation accuracy in the later stage.
	In addition, it should be noted that the algorithm will produce extremely small intermediate variables due to a large $\mu$, thus degrading the accuracy.
	Therefore, it is necessary to set an upper bound $\mu_{\mathrm max}$ for $\mu$.

	\subsection{Optimizing the Precoding Matrix ${\bf{F}}$}
	Upon replacing the OF of \eqref{PF1} with $f \left( {\bf{F}} \right)$ given in \eqref{f(F)}, the subproblem for the optimization of $\bf F$ is approximated as follows
	\begin{subequations}\label{PF3}
		\begin{alignat}{2}
			\max_{\bf{F}} \quad & f\left( {\bf{F}} \right) \\
			\mbox{s.t.}\quad
			&{\bf{F}} \in {{\cal S}_F}.
		\end{alignat}
	\end{subequations}
	The OF $f\left( {\bf{F}} \right)$ is continuous and concave but is still too complex to optimize directly,
	which motivates us to adopt the MM algorithm.
	The MM algorithm \cite{hunter2004tutorial,7547360} is widely used for resource allocation in wireless communication networks\cite{9090356,zhou2019intelligent,2019arXiv190409573Y,8741198}.
	We will use the MM algorithm to solve a series of more tractable surrogate problems satisfying several conditions, instead of the original one.
	Denote the optimal solution of the surrogate problem at the $n$th iteration by ${{\bf{F}}^n}$.
	The resulting sequence of ${{\bf{F}}^n}$ is guaranteed to converge to the KKT point of Problem \eqref{PF3}\cite{zhou2019intelligent},
	and the sequence of OF values $\left\{ {f\left( {{{\bf{F}}^1}} \right),f\left( {{{\bf{F}}^2}} \right), \ldots } \right\}$ must be monotonically non-decreasing.
	
	To describe the conditions that the OF of the surrogate problems must satisfy, we define $f'\left({\bf x}^{n};{\bf d}\right)$ as the directional derivative of $f\left({\bf x}^{n}\right)$, i.e.
	\[f'\left( {{{\bf{x}}^n};{\bf{d}}} \right) = \mathop {\lim }\limits_{\lambda  \to 0} \frac{{f\left( {{{\bf{x}}^n} + \lambda {\bf{d}}} \right) - f\left( {{{\bf{x}}^n}} \right)}}{\lambda }.\]
	The OF of the surrogate problem introduced at the $\left( t + 1 \right)$st iteration, denoted by $\tilde f\left( {{\bf{F}} | {{\bf{F}}^n}} \right)$, is said to minorize $f\left( {\bf{F}} \right)$ if\cite{7547360}
	
	\begin{enumerate}[({A}1)]
		\setlength{\itemsep}{0.7ex}
		
		\item $\tilde f\left( {{{\bf{F}}^n}| {{\bf{F}}^n}} \right) = f\left( {{{\bf{F}}^n}} \right),\forall {{\bf{F}}^n} \in {{\cal S}_F}$;
		
		\item $\tilde f\left( {{\bf{F}} | {\bf{F}}} \right) \le f\left( {\bf{F}} \right),\forall {\bf{F}},{{\bf{F}}^n} \in {{\cal S}_F}$;
		
		\item $\tilde f'\left( {{\bf{F}} | {{\bf{F}}^n};{\bf{d}}} \right){| _{{\bf{F}} = {{\bf{F}}^n}}} = f'\left( {{{\bf{F}}^n};{\bf{d}}} \right),\forall {\bf{d}} \ {\rm{ with }} \ {{\bf{F}}^n} + {\bf{d}} \in {{\cal S}_F}$;
		
		\item $\tilde f\left( {{\bf{F}}|{{\bf{F}}^n}} \right)$ is continuous in ${\bf{F}}$ and ${\bf{F}}^n$.
		
	\end{enumerate}
	To obtiain the surrogate problems, we introduce the following theorem:
	\begin{theorem}\label{theo_of_fF}
		For any feasible ${\bf{F}}$, $f\left( {\bf{F}} \right)$ is minorized with a quadratic function at solution ${\bf{F}}^n$ as follows
		\begin{equation}\label{tildefF}
			\tilde f\left( {{\bf{F}}|{{\bf{F}}^n}} \right) = {2{\mathop{\rm Re}\nolimits} \left\{ {{\mathop{\rm Tr}\nolimits} \left[ {{{\bf{V}}^{\rm H}}{\bf{F}}} \right]} \right\} + \alpha {\mathop{\rm Tr}\nolimits} \left[ {{{\bf{F}}^{\rm H}}{\bf{F}}} \right] + {\rm{cons}}F},
		\end{equation}
		In \eqref{tildefF}, ${\bf{V}}$ and ${\rm{cons}}F$ are respectively defined as
		\begin{subequations}
			\begin{alignat}{2}
				&{\bf{V}}\triangleq \sum\limits_{k \in {\cal K}} {{g_{{\rm{D}},k}}\left( {{{\bf{F}}^n}} \right)\left( {{{\bf{C}}_k} - {\bf{B}}_k^{\rm H}{{\bf{F}}^n}} \right)}  - \alpha {{\bf{F}}^n} ,\label{VarforF_V}\\
				&{\rm{cons}}F \triangleq f\left( {{{\bf{F}}^n}} \right) + \alpha {\mathop{\rm Tr}\nolimits} \left[ {{{\left( {{{\bf{F}}^n}} \right)}^{\rm H}}{{\bf{F}}^n}} \right]\notag\\
				&\quad - 2{\mathop{\rm Re}\nolimits} \left\{ {{\mathop{\rm Tr}\nolimits} \left[ \sum\limits_{k \in {\cal K}} {{g_{{\rm{D}},k}}\left( {{{\bf{F}}^n}} \right)\left( {{\bf{C}}_k^{\rm H} - {{\left( {{{\bf{F}}^n}} \right)}^{\rm H}}{{\bf{B}}_k}} \right){{\bf{F}}^n}} \right]} \right\}, \label{tildefF_consF}
			\end{alignat}
		\end{subequations}
		where
		\begin{subequations}
			\begin{alignat}{2}
				&{g_{{\rm{D}},k}}\left( {{{\bf{F}}^n}} \right)\triangleq \frac{{\exp \left\{ { - \mu {h_{{\rm{D}},k}}\left( {{{\bf{F}}^n}} \right)} \right\}}}{{\sum\limits_{k \in {\cal K}} {\exp \left\{ { - \mu {h_{{\rm{D}},k}}\left( {{{\bf{F}}^n}} \right)} \right\}} }},k \in {\cal K}\label{gDkFn},\\
				&\alpha \triangleq  - \mathop {\max }\limits_{k} \left\{ {{\rm tp1}_k} \right\} - 2\mu \mathop {\max }\limits_{k} \left\{ {{\rm tp2}_k} \right\} \label{VarforF_alpha},\\
				&{{\rm tp1}_k}\triangleq{{\omega _{{\rm{D}},k}}{w_{{\rm{D}},k}}u_{{\rm{D}},k}^ * {u_{{\rm{D}},k}}{\bf{h}}_{{\rm r},k}^{\rm H}{\bf{\Phi }}{{\bf{G}}_{\rm t}}{\bf{G}}_{\rm t}^{\rm H}{{\bf{\Phi }}^{\rm H}}{{\bf{h}}_{{\rm r},k}}} \label{tp1k},\\
				&{{\rm tp2}_k}\triangleq{{P_{max}}{{\rm {tp1}}^2_k} + \left\| {{{\bf{C}}_k}} \right\|_F^2 + 2\sqrt {{P_{max}}} {{\left\| {{{\bf{B}}_k}{{\bf{C}}_k}} \right\|}_F}}.\label{VarforF_tp2k}
			\end{alignat}
		\end{subequations}
		
		\emph{Proof:} Please refer to Appendix \ref{appfF}.
	\end{theorem}
	
	We can formulate the surrogate problem for solving ${\bf{F}}$ at each iteration by replacing the OF of Problem \eqref{PF3} with \eqref{tildefF}, as follows
	\begin{subequations}\label{PF4}
		\begin{alignat}{2}
			\max_{{\bf{F}}} \quad &  {2{\mathop{\rm Re}\nolimits} \left\{ {{\mathop{\rm Tr}\nolimits} \left[ {{{\bf{V}}^{\rm H}}{\bf{F}}} \right]} \right\} + \alpha {\mathop{\rm Tr}\nolimits} \left[ {{{\bf{F}}^{\rm H}}{\bf{F}}} \right] + {\rm{cons}}F} \\
			\mbox{s.t.}\quad
			&{\bf{F}} \in {{\cal S}_{F}}.
		\end{alignat}
	\end{subequations}
	The optimal closed-form solution of Problem \eqref{PF4} can be obtained using the Lagrangian multiplier method.
	Introducing the Lagrange multiplier $\zeta$, the Lagrangian function is written as
	\begin{align}
		{\mathcal L}\left( {{\bf{F}},\zeta } \right) &= 2{\mathop{\rm Re}\nolimits} \left\{ {{\mathop{\rm Tr}\nolimits} \left[ {{{\bf{V}}^{\rm H}}{\bf{F}}} \right]} \right\} + \alpha {\mathop{\rm Tr}\nolimits} \left[ {{{\bf{F}}^{\rm H}}{\bf{F}}} \right] \notag \\
		&\quad + {\rm{cons}}F - \zeta \left( {{\mathop{\rm Tr}\nolimits} \left[ {{{\bf{F}}^{\rm H}}{\bf{F}}} \right] - {P_{\rm{max}}}} \right).
	\end{align}
	
	Setting the first-order derivative of ${\cal L}\left( {{\bf{F}},\zeta } \right)$ w.r.t. $\bf F$ to zero, we can obtain the solution of $\bf F$ as follows
	\begin{equation}
		{\bf{F}} = \frac{{\bf{V}}}{{ \zeta - \alpha}}.
	\end{equation}
	Given the power constraint ${\rm Tr} \left[ {{{\mathbf{F}}^{\rm H}}{\mathbf{F}}} \right] \le {P_{\rm{max}}}$, it follows that
	\begin{equation}\label{pow in V}
		\frac{{{\mathop{\rm Tr}\nolimits} \left[ {{{\bf{V}}^{\rm H}}{\bf{V}}} \right]}}{{{{\left( {\zeta  - \alpha} \right)}^2}}} \le {P_{\rm{max}}}.
	\end{equation}
	The left hand side of \eqref{pow in V} is a decreasing function w.r.t $\zeta$.
	As a result, we obtain the optimal solution of $\bf F$ at the $n$th iteration as follows
	\begin{equation}\label{Fn+1}
		{{\bf F}^{n+1}}=\begin{cases}
			-{\bf V}/\alpha, & \text{if}\ \eqref{pow in V} \text{ holds when } \zeta=0 ;\\
			-\sqrt {{P_{\rm{max}}}/{\mathop{\rm Tr}\nolimits} \left[ {{{\bf{V}}^{\rm H}}{\bf{V}}} \right]} {\bf{V}}, & \text{otherwise}.
		\end{cases}
	\end{equation}
	
	\subsection{Optimizing the Reflection Coefficient Vector $\bm \phi$}
	Replacing the OF of \eqref{Pphi1} with $f\left( {\bm{\phi}} \right)$ given in \eqref{f(phi)}, the approximated subproblem for the reflection coefficient vector $\bm \phi$ is given as follows
	\begin{subequations}\label{Pphi3}
		\begin{alignat}{2}
			\max_{\bm{\phi}} \quad & f\left( {\bm{\phi}}  \right) \\
			\mbox{s.t.}\quad
			&{\bm{\phi}} \in {{\cal S}_{\phi}}\label{Pphi3b}.
		\end{alignat}
	\end{subequations}
	Similar to the process of optimizing $\bf F$ in the previous subsection, we adopt the MM algorithm framework.
	Note that constraint \eqref{Pphi3b} is non-convex.
	To guarantee convergence, the conditions of the minorizing function $\tilde f\left( {{{\bm{\phi}}}| {{\bm{\phi}}}^n} \right)$ should be modified as follows
	\cite{2691647620070801,12133060520170201}
	
	\begin{enumerate}[({B}1)]
		\setlength{\itemsep}{0.7ex}
		\item $\tilde f\left( {{{\bm{\phi}}^n}| {{\bm{\phi}}}^n} \right) = f\left( {{{\bm{\phi}}^n}} \right),\forall {{\bm{\phi}}^n} \in {{\cal S}_{\phi}}$;
		
		\item $\tilde f\left( {{\bm{\phi}} | {\bm{\phi}}^n} \right) \le f\left( {\bm{\phi}} \right),\forall {\bm{\phi}},{{\bm{\phi}}^n} \in {{\cal S}_{\phi}}$;
		
		\item $\tilde f'\left( {{\bm{\phi}} | {{\bm{\phi}}^n};{\bf{d}}} \right){| _{{\bm{\phi}} = {{\bm{\phi}}^n}}} = f'\left( {{{\bm{\phi}}^n};{\bf{d}}} \right),\forall {\bf{d}} \in {{\cal J}}_{{\cal S}_{\phi}} \left({\bm{\phi}}\right)$;
		
		\item $\tilde f\left( {{\bm{\phi}}|{{\bm{\phi}}^n}} \right)$ is continuous in ${\bm{\phi}}$ and ${\bm{\phi}}^n$.
		
	\end{enumerate}
	where ${{\cal J}}_{{\cal S}_{\phi}} \left({\bm{\phi}}\right)$ is the Boulingand tangent cone of ${{\cal S}_{\phi}}$.
	A feasible $\tilde f\left( {{\bm{\phi}}|{{\bm{\phi}}^n}} \right)$ can be constructed as shown in the following theorem:
	\begin{theorem}\label{theo_of_fphi}
		For any feasible ${\bm{\phi}}$, $f\left( {\bm{\phi}} \right)$ is minorized with the following function:
		\begin{equation}\label{tildefphi}
			\tilde f\left( {{\bm{\phi}} |{{\bm{\phi}} ^n}} \right) = 2{\mathop{\rm Re}\nolimits} \left\{ {{{\bf{v}}^{\rm H}}{\bm{\phi}} } \right\} + {\rm{cons}}\phi,
		\end{equation}
		In \eqref{tildefphi}, ${\bf{v}}$ and ${\rm{cons}}\phi$ are respectively defined as
		\begin{subequations}
			\begin{alignat}{2}
				&{\bf{v}} \triangleq {\bf{d}} - \beta {{\bm{\phi}} ^n}\label{Varforphi_v},\\
				&{\rm{cons}}\phi  \triangleq f\left( {{{\bm{\phi}} ^n}} \right) + 2M\beta  - 2{\mathop{\rm Re}\nolimits} \left\{ {{{\bf{d}}^{\rm H}}{{\bm{\phi}} ^n}} \right\}\label{Varforphi_consphi},
			\end{alignat}
		\end{subequations}
		where
		\begin{subequations}
			\begin{alignat}{2}
				&{\bf{d}} \triangleq \sum\limits_{l \in {\cal L}} {\sum\limits_{k \in {\cal K}} {{g_{l,k}}\left( {\bm{\phi}}^n  \right)\left( {{{\bf{a}}_{l,k}} - {\bf{A}}_{l,k}^{\rm H}{{\bm{\phi}} ^n}} \right)} } ,\\
				&{g_{l,k}}\left( {{{\bm{\phi}} ^n}} \right) \triangleq \frac{{\exp \left\{ { - \mu {h_{l,k}}\left( {{{\bm{\phi}} ^n}} \right)} \right\}}}{{\sum\limits_{l \in {\cal L}} {\sum\limits_{k \in {\cal K}} {\exp \left\{ { - \mu {h_{l,k}}\left( {{{\bm{\phi}} ^n}} \right)} \right\}} } }},l \in {\cal L},k \in {\cal K},\label{glkphin}\\
				&\beta  \triangleq- 2\mu \mathop {\max }\limits_{l ,k } \left\{ {\left\| {{{\bf{a}}_{l,k}}} \right\|_2^2 + M{\lambda _{\max }}\left( {{{\bf{A}}_{l,k}}{\bf{A}}_{l,k}^{\rm H}} \right) + 2{{\left\| {{{\bf{A}}_{l,k}}{{\bf{a}}_{l,k}}} \right\|}_1}} \right\}\notag\\
				& \qquad - \mathop {\max }\limits_{l,k} \left\{ {{\lambda _{\max }}\left( {{{\bf{A}}_{l,k}}} \right)} \right\}.\label{Varforphi_beta}
			\end{alignat}
		\end{subequations}
		
		\emph{Proof:} Please refer to Appendix \ref{appfphi}.
		
	\end{theorem}
	
	The surrogate problems of $\bm{\phi}$ at each iteration with closed-form solutions is formulated by replacing the OF of Problem \eqref{Pphi3} with \eqref{tildefphi}, as follows
	\begin{subequations}\label{Pphi4}
		\begin{alignat}{2}
			\max_{\bm{\phi}} \quad & 2{\mathop{\rm Re}\nolimits} \left\{ {{{\bf{v}}^{\rm H}}{\bm{\phi}} } \right\} + {\rm{cons}}\phi \\
			\mbox{s.t.}\quad
			&{\bm{\phi}} \in {{\cal S}_{\phi}}.
		\end{alignat}
	\end{subequations}
	The optimal solution of $\bm \phi$ at the $n$th iteration is given by
	\begin{equation}\label{phin+1}
		{{\bm{\phi}} ^{n + 1}} = \exp \left\{ {j\angle {\bf{v}}} \right\},
	\end{equation}
	where $\angle \left(  \cdot  \right)$ and $\exp \left\{  \cdot  \right\}$ are element-wise operations as before.
	
	\subsection{Algorithm Development}
	In theory, by adopting the MM method to solve the subproblems \eqref{PF4} and \eqref{Pphi4} instead of solving Problem \eqref{P2} directly,
	the precoding matrix $\bf F$ and the reflection coefficient vector $\bm \phi$ can be optimized at a lower computational cost.
	However, the convergence speed of the proposed MM algorithm is limited by the tightness of the minorizing functions $\tilde f\left( {{\bf{F}}|{{\bf{F}}^n}} \right)$ and $\tilde f\left( {{\bm{\phi}} |{{\bm{\phi}} ^n}} \right)$, which is mainly determined by $\alpha$ in \eqref{VarforF_alpha} and $\beta$ in \eqref{Varforphi_beta}.
	Although the MM algorithm requires little computation per iteration,
	the large number of iterations required for convergence may lead to a long total operation time.
	Therefore, we introduce SQUAREM\cite{varadhan2008simple} theory to accelerate the convergence of the proposed MM algorithm.
	Specifically, the number of MM iterations required at each update of $\bf F$ or $\bm \phi$ is reduced to 2.
	
	\subsubsection{BCD-MM  algorithm}
	\begin{algorithm}[t]
		\caption{BCD-MM algorithm}
		\label{BCD-MM}
		\begin{algorithmic}[1]
			\State Initialize iteration number $n=1$ and feasible ${\bf F}^1$ and ${\bm \phi}^1$.
			Calculate ${{\rm Obj}} \left( {{{\bf{F}}^1},{{\bm \phi} ^1}} \right)$.
			Set $\mu$, $\mu_{\mathrm max}$, $\iota$, maximum number of iterations $n_{\rm max}$ and error tolerance $\varepsilon_{\rm e}$;
			\State Given ${\bf F}^{n}$ and ${\bm \phi}^n$, calculate the optimal decoding variables ${\mathcal U}_{\rm{D}}^{n+1}$ in \eqref{optuD} and the optimal linear receivers ${\mathcal U}_{\rm{U}}^{n+1}$ in \eqref{optuU};
			\State Given ${\bf F}^{n}$, ${\bm \phi}^n$, ${\mathcal U}_{\rm{D}}^{n+1}$ and ${\mathcal U}_{\rm{U}}^{n+1}$, calculate the optimal auxiliary variables ${\mathcal W}_{\rm{D}}^{n+1}$ and ${\mathcal W}_{\rm{U}}^{n+1}$ in \eqref{optw};
			\State Calculate ${\bf F}_1=\mathfrak{M}_F \left({\bf F}^{n}\right)$ and ${\bf F}_2=\mathfrak{M}_F \left({\bf F}_1\right)$;\label{iter_map_F}
			\State Calculate ${\bf Q}_1={\bf F}_1-{\bf F}^{n}$ and ${\bf Q}_2={\bf F}_2-{\bf F}_1-{\bf Q}_1$;
			\State Calculate step factor $\varpi  =  - \frac{{{{\left\| {{{\bf{Q}}_1}} \right\|}_F}}}{{{{\left\| {{{\bf{Q}}_2}} \right\|}_F}}}$;
			\State Calculate ${{\bf{F}}^{n + 1}} = {{\bf{F}}^n} - 2\varpi {{\bf{Q}}_1} + {\varpi ^2}{{\bf{Q}}_2}$.\label{update_F_BCDMM}
			\State If ${{\bf{F}}^{n + 1}} \notin {{\mathcal S}_F}$, scale ${{\bf{F}}^{n + 1}}\leftarrow \frac{{\sqrt {{P_{{\rm{max}}}}} }}{{\left\| {{{\bf{F}}^{n + 1}}} \right\|}}{{\bf{F}}^{n + 1}}$;
			\State If $f\left( {{\bf{F}}^{n+1}} \right)|_{ {\bm \phi}={{\bm \phi} ^n}} <f\left( {{\bf{F}}^{n}} \right)|_{ {\bm \phi}={{\bm \phi} ^n}}$, set $\varpi \leftarrow \left( {\varpi  - 1} \right)/2$ and go to step \ref{update_F_BCDMM};\label{update_F_BCDMM_end}
			\State Calculate ${\bm \phi}_1=\mathfrak{M}_{\phi} \left({\bm \phi}^{n}\right)$ and ${\bm \phi}_2=\mathfrak{M}_{\phi} \left({\bm \phi}_1\right)$;\label{iter_map_phi}\label{update_phi_BCDMM_begin}
			\State Calculate ${\bf q}_1={\bm \phi}_1-{\bm \phi}^{n}$ and ${\bf q}_2={\bm \phi}_2-{\bm \phi}_1-{\bf q}_1$;
			\State Calculate step factor $\varpi  =  - \frac{{{{\left\| {{{\bf{q}}_1}} \right\|}_F}}}{{{{\left\| {{{\bf{q}}_2}} \right\|}_F}}}$;
			\State Calculate ${{\bm \phi}^{n + 1}} = \exp \left\{ {\angle \left( {{{\bm \phi}^n} - 2\varpi {{\bf{q}}_1} + {\varpi ^2}{{\bf{q}}_2}} \right)} \right\}$;\label{update_phi_BCDMM}
			\State If $f\left( \boldsymbol{\phi }^{n+1} \right) |_{\mathbf{F}=\mathbf{F}^{n+1}}<f\left( \boldsymbol{\phi }^n \right) |_{\mathbf{F}=\mathbf{F}^{n+1}}$, set $\varpi \leftarrow \left( {\varpi  - 1} \right)/2$ and go to step \ref{update_phi_BCDMM};\label{update_phi_BCDMM_end}
			\State Set $\mu \leftarrow \max \left(\mu^\iota,\mu_{\mathrm max}\right)$;\label{increse_mu_BCMMM}
			\State If $\left|{{\rm Obj}} \left( {{{\bf{F}}^{n+1}},{{\bm \phi} ^{n+1}}} \right)-{{\rm Obj}} \left( {{{\bf{F}}^{n}},{{\bm \phi} ^{n}}} \right)\right|/{{\rm Obj}} \left( {{{\bf{F}}^{n}},{{\bm \phi} ^{n}}} \right)<\varepsilon_{\rm e}$ or $n\ge n_{\rm max} $, terminate.
			Otherwise, set $n\leftarrow n+1$ and go to step 2.
		\end{algorithmic}
	\end{algorithm}
	The accelerated version of our proposed algorithm referred to as BCD-MM, is detailed in Algorithm \ref{BCD-MM},
	where the OF of Problem \eqref{P1} evaluated at ${{\bf{F}}^n}$ and ${{\bm \phi} ^n}$ is denoted as ${{\rm Obj}} \left( {{{\bf{F}}^n},{{\bm \phi} ^n}} \right)$,
	and the original MM iteration rules of $\bf F$ given in \eqref{Fn+1} and those of $\bm \phi$ given in \eqref{phin+1} are denoted as the nonlinear fixed-point iteration maps $\mathfrak{M}_F\left(\cdot\right)$ and $\mathfrak{M}_{\phi}\left(\cdot\right)$, respectively.
	As shown in step \ref{increse_mu_BCMMM}, we propose to define an adjustment factor $\iota$ to gradually increase $\mu$ to $\mu_{\mathrm max}$.
	
	The MM method yields monotonically non-decreasing OF values for \eqref{PF3} and \eqref{Pphi3},
	i.e. $ f\left( {{{\bf{F}}^n}} \right) \leq  f\left( {{{\bf{F}}_1}} \right) \leq  f\left( {{{\bf{F}}_2}} \right)$ and $ f\left( {{{\bm{\phi}}^n}} \right) \leq  f\left( {{{\bm{\phi}}_1}} \right) \leq  f\left( {{{\bm{\phi}}_2}} \right)$.
	Both steps \ref{update_F_BCDMM_end} and \ref{update_phi_BCDMM_end} ensure that the value of the OF in Problem \eqref{P1} is non-decreasing.
	Additionally, the value of the OF must have an upper bound,
	due to the limitations on the maximum transmit power $P_{\rm max}$ and the number of reflection elements $M$.
	Hence, Algorithm \ref{BCD-MM} is guaranteed to converge.
	
	Note that the KKT optimality of the converged solution of MM algorithm has been proved and verified widely in existing literatures,
	such as \cite{zhou2019intelligent}, \cite{9090356} and \cite{7547360}.
	Hence, the converged solution $\left\{\mathbf{F}^{\star},\boldsymbol{\phi }^{\star}\right\}$ generated by Algorithm \ref{BCD-MM} satisfies the KKT conditions of problems \eqref{PF3} and \eqref{Pphi3}.
	When the first equality in \eqref{smoothchar_F} and that in \eqref{smoothchar_phi} hold,
	problems \eqref{PF3} and \eqref{Pphi3} are respectively equivalent to \eqref{PF2} and \eqref{Pphi2}.
	Then, it can be readily verified that $\left\{\mathbf{F}^{\star},\boldsymbol{\phi }^{\star}\right\}$ satisfies the KKT conditions of Problem \eqref{P1}.
	In fact, the approximations in \eqref{smoothchar_F} and \eqref{smoothchar_phi} are very tight when smoothing parameter $\mu$ is sufficiently large.
	That is, the converged solution of Algorithm \ref{BCD-MM} is very close to a KKT point of Problem \eqref{P1}.
	Moreover, by gradually increasing $\mu$ to reduce the approximation gap,
	Algorithm \ref{BCD-MM} actually has a relatively high probability of converging to a good locally optimal solution of Problem \eqref{Pphi2}.
	
	\subsubsection{Complexity Analysis}\label{comp_analy_BCD-MM}
	First, as discussed in \ref{comp_analy_BCD}, the complexity of computing ${\mathcal U}_{\rm D}$, ${\mathcal U}_{\rm U}$, ${\mathcal W}_{\rm D}$ and ${\mathcal W}_{\rm U}$ is of order ${\cal O}\left(K^2\left(M^2+N_{\rm t} M+N_{\rm r} M\right)+KM^3\right)$.

	Second, we analyze the computational complexity of solving Problem \eqref{PF1} and \eqref{Pphi2} with the proposed MM algorithm.
	The computational complexity of optimizing $\bf F$ lies mainly in the calculation of $\bf V$ in \eqref{VarforF_V} and $\alpha$ in \eqref{VarforF_alpha},
	whose complexity in turn depends on ${g_{{\rm{D}},k}}$ in \eqref{gDkFn} and ${\rm tp2}_k$ in \eqref{VarforF_tp2k}, respectively.
	Since the $K$ values of ${h_{{\rm{D}},k}}\left( {{{\bf{F}}^n}} \right)$ are repeated in every ${g_{{\rm{D}},k}}\left( {{{\bf{F}}^n}} \right)$, the complexity of computing ${g_{{\rm{D}},k}}$ is ${\cal O}\left(K\left(N_{\rm t} M^2+K^2 N_{\rm t}+K N_{\rm t}^2\right)\right)$.
	The complexity for each ${\rm tp2}_k$ is ${\cal O}\left(K^2 N_{\rm t}+K N_{\rm t}^2\right)$,
	so the complexity order for $\alpha$ is ${\cal O}\left(K\left(K^2 N_{\rm t}+K N_{\rm t}^2\right)\right)$.
	Recall that to calculate ${\bf F}^{n+1}$ and ${\bm \phi}^{n+1}$, only two MM iterations are required in each BCD iteration.
	Hence, the complexity of calculating ${\bf F}^{n+1}$ is given by ${\cal O}\left(K\left(N_{\rm t} M^2+K^2 N_{\rm t}+K N_{\rm t}^2\right)\right)$.
	The calculation of ${g_{l,k}}\left( {{{\bm{\phi}} ^n}} \right)$ in \eqref{glkphin} and $\beta$ in \eqref{Varforphi_beta} comprises the main complexity of calculating $ {{{\bm{\phi}} ^{n+1}}}$.
	The order of complexity for each ${h_{l,k}}\left( {{{\bm{\phi}} ^n}} \right)$ is ${\cal O}\left(K M^2\right)$,
	and thus that for ${g_{l,k}}\left( {{{\bm{\phi}} ^n}} \right)$ is ${\cal O}\left(K^2 M^2\right)$.
	Additionally, the computational complexity of calculating the maximum eigenvalues of ${\bf A}_{l,k}$ and ${{{\bf{A}}_{l,k}}{\bf{A}}_{l,k}^{\rm H}}$ is of order ${\cal O}\left(M^3\right)$.
	Thus the computational complexity of calculating $\beta$ in \eqref{Varforphi_beta} is of order ${\cal O}\left(KM^3\right)$,
	and that for ${\bm \phi}^{n+1}$ is ${\cal O}\left(K^2 M^2+KM^3\right)$.
	
	Finally, the overall complexity of Algorithm \ref{BCD-MM} is of order
	\begin{align}\label{complex_Alg2}
		{\cal C}_{\rm {Alg. \ref{BCD-MM}}}&={\cal O}\left( {K^2}{N_{\rm{t}}}M  + {K^2}{N_{\rm{r}}}M + {K^2}N_{\rm{t}}^2 + {K^3}{N_{\rm{t}}}\right)\notag \\
		&\quad + {\cal O}\left( {K{M^3} + K{N_{\rm{t}}}{M^2} + {K^2}{M^2}} \right).
	\end{align}
	Clearly, application of the MM method greatly reduces the computational load of the algorithm.
	
	\section{Simulation Results} \label{SIMULATION RESULTS}
	In this section, extensive simulation results are presented to verify the performance of the proposed multiuser IRS-aided FD two-way communication system.
	
	\begin{figure}
		\centering
		\includegraphics[scale=0.51]{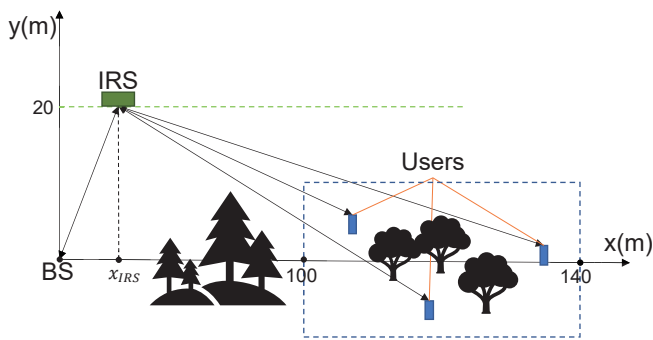}
		\caption{The simulated IRS-aided FD two-way multiuser communication scenario.}
		\label{Figsimulmodel}
	\end{figure}
	
	\subsection{Simulation Setup}
	Fig. \ref{Figsimulmodel} shows the horizontal plane of the schematic system model for our simulated network.
	As shown in the figure, we consider a system with $K=3$ users,
	whose coordinates are generated uniformly and randomly in a rectangular region centered at $\left(120,0\right)$ with length 40 m and width 20 m.
	The coordinates of the BS and the IRS are $\left(0,0\right)$ and $\left(x_{\rm IRS},20\right)$, respectively, where the default value of $x_{\rm IRS}$ is 10.
	We assume that the height of the BS, the IRS, and the users are $30 \;\text{m}$, $10 \;\text{m}$, and $1.5 \;\text{m}$ \cite{9090356}, respectively.
	
	The path loss is taken to be -30 dB at a reference distance of 1 m.
	The path loss exponents of the links between the BS and the IRS as well as those of the links between the IRS and the users are denoted by $\alpha_{\rm{BI}}$ and $\alpha_{\rm{IU}}$, respectively.
	As we stated in Section \ref{SYSTEM MODEL AND PROBLEM FORMULATION}, there is no direct link between the BS and the users.
	On the contrary, through proper site selection, the transmission environment for the IRS-provided link can be nearly free-space.
	Hence, we set $\alpha_{\rm{BI}}=\alpha_{\rm{IU}}=\alpha_{\rm{IRS}}=2.2$ \cite{9090356}.
	Then, the large-scale path loss in dB is modeled by
	\begin{equation}
		{\rm{PL}} =-30  - 10\alpha {\log _{10}}d,
	\end{equation}
	where $d$ is the link distance beyond the 1 m reference.
	The small-scale fading is assumed to be Rician distributed, modeled by
	\begin{equation}
		{\bf{\tilde G}} = \sqrt {\frac{\kappa }{{\kappa  + 1}}} {{\bf{\tilde G}}^{\rm{LoS}}} + \sqrt {\frac{1}{{\kappa  + 1}}} {{\bf{\tilde G}}^{\rm{NLoS}}},
	\end{equation}
	where $\kappa$ is the Rician factor, ${{\bf{\tilde G}}^{\rm{LoS}}}$ and ${{\bf{\tilde G}}^{\rm{NLoS}}}$ are the LoS and the NLoS components, respectively.
	${{\bf{\tilde G}}^{\rm{NLoS}}}$ is drawn from a Rayleigh distribution, and ${{\bf{\tilde G}}^{\rm{LoS}}}$ is modeled as
	\begin{equation}\label{G_LOS}
		{{\bf{\tilde G}}^{\rm{LoS}}}={{\bf{c}}_r}\left( {{\vartheta ^{\rm{AoA}}}} \right){{\bf{c}}^{\rm H}_t}\left( {{\vartheta ^{\rm{AoD}}}} \right).
	\end{equation}
	In \eqref{G_LOS}, ${{\bf{c}}_r}\left( {{\vartheta ^{\rm{AoA}}}} \right)$ and ${{\bf{c}}_t}\left( {{\vartheta ^{\rm{AoD}}}} \right)$ are respectively given by
	\begin{subequations}
		\begin{alignat}{2}
			&{{\bf{c}}_r}\left( {{\vartheta ^{\rm{AoA}}}} \right) = {\left[ {1,{e^{j\pi \sin {\vartheta ^{\rm{AoA}}}}}, \ldots ,{e^{j\pi \left( {{W_r} - 1} \right)\sin {\vartheta ^{\rm{AoA}}}}}} \right]^{\rm T}},\\
			&{{\bf{c}}_t}\left( {{\vartheta ^{\rm{AoD}}}} \right) = {\left[ {1,{e^{j\pi \sin {\vartheta ^{\rm{AoD}}}}}, \ldots ,{e^{j\pi \left( {{W_t} - 1} \right)\sin {\vartheta ^{\rm{AoD}}}}}} \right]^{\rm T}},
		\end{alignat}
	\end{subequations}
	where $W_{\rm r}$ and $W_{\rm t}$ denote the number of antennas/elements at the receiver side and transmitter side, respectively, ${{\vartheta ^{\rm{AoA}}}}$ and ${{\vartheta ^{\rm{AoD}}}}$ are the angle of arrival and departure, respectively.
	In the simulations, we independently and randomly generate ${{\vartheta ^{\rm{AoA}}}}$ and ${{\vartheta ^{\rm{AoD}}}}$ in the range of $\left[0,2\pi\right]$.
	For simplicity, we set $\sigma _{{\rm{U}}}^2=1.1 \sigma _{\rm B}^2$ and $\sigma _{{\rm{D}},k}^2=1.1 \sigma _{k}^2, \forall k$.
	Unless otherwise stated, the other parameters are set as follows:
	Channel bandwidth 10 MHz, Rician factor $\kappa=3$, noise power density -174 dBm/Hz, SI coefficient $\rho_{\rm S}=1$,
	weighting factors $\omega_{l,k}=1,\forall l,k$,
	user transmit power $P_{k}=50$ mW, $\forall k$,
	number of BS antennas $N_{\rm t}=N_{\rm r}=4$,
	maximum BS transmit power $P_{\rm {max}}=1\text{ W}$,
	number of IRS reflection elements $M=16$,
	x-coordinate of IRS $x_{\rm IRS}=10$ m,
	initial smoothing parameter $\mu=5$, adjusting factor $\iota=1.02$, upper bound $\mu_{\mathrm max}=500$,
	error tolerance $\varepsilon_{\rm e}=10^{-6}$.
	The following results are obtained by averaging over 200 independent channel realizations.
	The reflection coefficient vector $\bm \phi$ is initialized by uniformly and randomly selecting the phase shift of each reflection element in $\left[0,2\pi\right]$.
	The precoding matrix $\bf F$ is initialized by extracting the real and imaginary parts of each element of $\bf F$ from an independent Gaussian distribution, and then scaling $\bf F$ to satisfy the equality in \eqref{pow_constraint}.
	
	\subsection{Baseline Schemes}
	In our simulation, Problem \eqref{PF2} and the relaxed version of Problem \eqref{Pphi2} in Algorithm \ref{SCOP-based BCD} are solved using the MOSEK solver \cite{Themosekoptimizationtoolbox}.
	In the remainder of this section, we denote the proposed Algorithm \ref{SCOP-based BCD} by \textbf{BCD-SOCP}, and Algorithm \ref{BCD-MM} by \textbf{BCD-MM}.
	In order to analyze the performance of our proposed algorithms, we consider three baseline schemes:
	\begin{enumerate}
		\item Note that MOSEK solver can optimally solve the SOCP \eqref{PF2}.
		To compare the performance of our proposed algorithms in solving quasi-SOCP, we design a benchmark algorithm \textbf{SOCP+MM} by replacing steps \ref{iter_map_F} to \ref{update_F_BCDMM_end} of Algorithm \ref{BCD-MM} with step \ref{update_F_BCD} of Algorithm \ref{SCOP-based BCD}.
		
		\item To analyse the benefits of jointly optimizing the precoding matrix and the reflection coefficient vector, we consider the schemes in which only the former is optimized.
		Specifically, the steps that update the value of $\bm \phi$ are skipped.
		We refer to implementing Algorithm \ref{SCOP-based BCD} with random phase as \textbf{BCD-SOCP, Rand}.
		A similar definition holds for \textbf{BCD-MM, Rand}.
		\item Since an IRS with arbitrarily tunable phase shifts is difficult to implement, we consider a more practical scenario involving 2-bit control of each IRS element (e.g., 4 possible phase shifts per element).
		Specifically, each element of the optimal reflection coefficient vector ${\bm \phi}^{\rm opt}$ obtained by \textbf{BCD-SOCP}, \textbf{BCD-MM} or \textbf{SOCP+MM} is converted to the following quantized value:
		\begin{equation}
			\phi _m^{{\rm{2-bit}}} = \exp \left\{ {\arg \min_\theta \left| {\angle \phi _m^{{\rm{opt}}} - \theta } \right|} \right\},m = 1, \ldots ,M,
		\end{equation}
		where $\theta  \in \left\{ {0,\frac{\pi }{2},\pi ,\frac{{3\pi }}{2}} \right\}$.
		The corresponding $\bf F$ is then updated.
		The resulting algorithms are denoted as \textbf{BCD-SOCP, 2 bit}, \textbf{BCD-MM, 2 bit} and \textbf{SOCP+MM, 2 bit}, respectively.
	\end{enumerate}

	\subsection{Convergence of Proposed Algorithm}
	
	
	\begin{figure}
		\centering
		\subfigure[Achievable WMR versus the number of iterations]{
			\label{fig:conv-iter} 
			\includegraphics[width=1\linewidth]{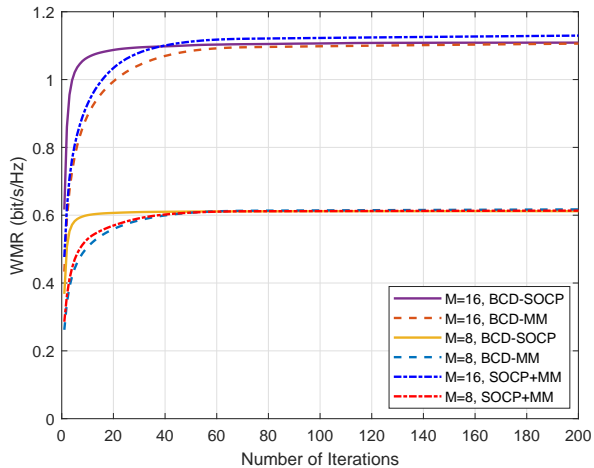}}
		\subfigure[Achievable WMR versus CPU time]{
			\label{fig:conv-time} 
			\includegraphics[width=1\linewidth]{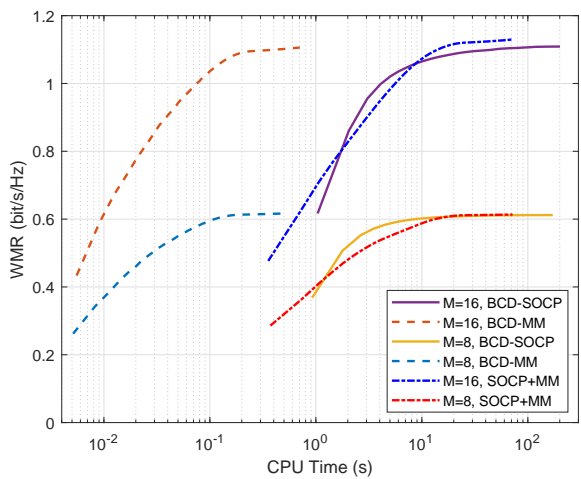}}
		\caption{Convergence behaviour of proposed algorithms for $M=\left[8,16\right]$}
		\label{fig:convergence} 
	\end{figure}
	

	Fig. \ref{fig:convergence} plots the WMR versus the number of iterations and the CPU time for $M=8 \; \text{and}\; 16$,
	illustrating the convergence behaviour of our proposed algorithms and the benchmark algorithm.
	200 iterations of each algorithm are performed in each trial.
	We see that all the algorithms converge within 80 iterations, which confirms their high efficiency.
	The converged WMR of \textbf{BCD-SOCP} and \textbf{BCD-MM} are basically the same, and both are slightly lower than that of \textbf{SOCP+MM}, which shows the accuracy of MOSEK in solving SOCP and MM algorithm's advantages in finding globally optimal solution of quasi-SOCP.
	However, due to its advantage in computational complexity, \textbf{BCD-MM} converges much faster in terms of CPU time.
	Additionally, it is interesting to observe that even when the number of reflection coefficients doubles,
	its convergence speed in terms of both number of iterations and CPU time does not increase significantly.
	The explanation can be found in the updating strategy for the smoothing factor and the computational complexity of the algorithms.
	On the one hand, the convergence speed of Algorithm \ref{BCD-MM} mainly depends on the approximation level of the surrogate functions in the MM iterations, which is mainly controlled by $\mu$, whose rate of increase is set to gradually accelerate.
	On the other hand, the quadratic and cubic terms in $M$ only account for less than half of the seven terms in the expression for the computational complexity of Algorithm \ref{BCD-MM} in \eqref{complex_Alg2}.
	This indicates that our proposed Algorithm \ref{BCD-MM} will maintain good convergence performance and relatively low complexity even for the case of large $M$.

	\subsection{Impact of the IRS Location}
	\begin{figure}
		\centering
		\subfigure[Achievable WMR versus $x_{\rm {IRS}}$ for $\rho_S=1$.]{
			\label{fig:xIRS-SI1} 
			\includegraphics[width=1\linewidth]{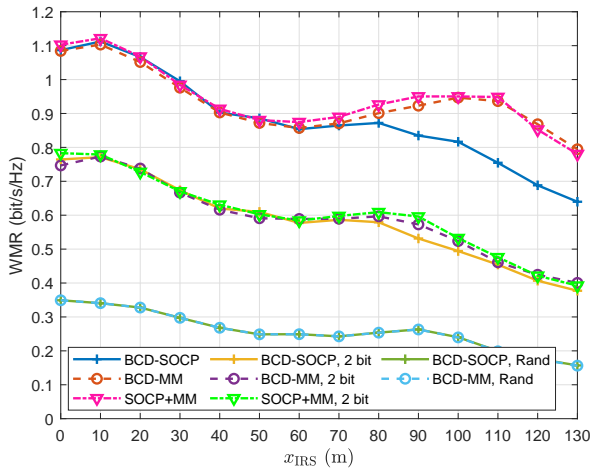}}
		\subfigure[Achievable WMR versus $x_{\rm {IRS}}$ for $\rho_S=0.1$.]{
			\label{fig:xIRS-SI0.1} 
			\includegraphics[width=1\linewidth]{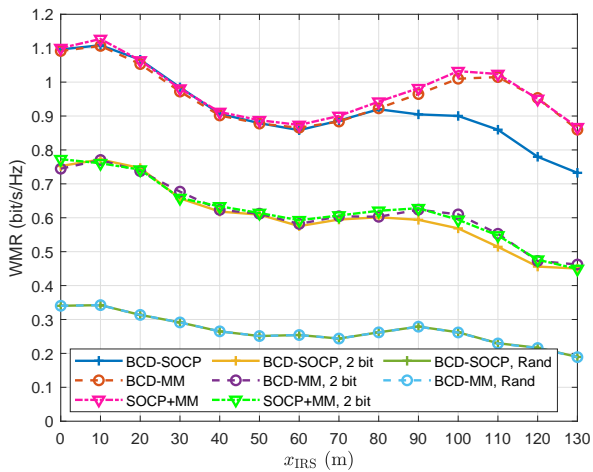}}
		\caption{Impact of the IRS location $x_{\rm {IRS}}$ and SI coefficient $\rho_S$.}
		\label{fig:impact_loca_rho} 
	\end{figure}

	In order to provide engineering guidance for IRS site selection in practical communication systems, we investigate the effect of IRS location on the achievable WMR.
	By moving the IRS along the dotted line in Fig. \ref{Figsimulmodel} from $x_{\rm {IRS}}=0$ to $x_{\rm {IRS}}=130$,
	Fig. \ref{fig:xIRS-SI1} and Fig. \ref{fig:xIRS-SI0.1} illustrate the impact of IRS location on the achievable WMR for two cases of the SI coefficient $\rho_S=1$ and $\rho_S=0.1$, respectively.
	We can first draw a preliminary conclusion from these figures that for all eight schemes, IRS deployments nearer the BS improve the WMR.
	Second, recall that the x-coordinate of the users is distributed independently and uniformly between 100 and 140 in our simulation.
	Let us loosely name the point (120,0) as the \emph{user central point}, and name the space on the left and right side of $x=60$ as the \emph{BS side} and the \emph{user side}, respectively.
	Then, it can be observed that there are always two peaks in the achievable WMR for the various schemes, one on the BS side and one on the user side.
	Due to the increase in path loss, the achievable WMR decreases as expected when $x_{\rm {IRS}}$ is too small or too large.
	Furthermore, the valley value of the WMR that occurs when $x_{\rm {IRS}}\approx60$ may also be explained by path loss.
	We can approximate the large-scale channel gain as follows
	\begin{align}\label{l-s_chan_gain}
		{\rm{P}}{{\rm{L}}_{{\rm{IRS}}}} =  - 60 - 10\alpha {\log _{10}}\left( {{x_{{\rm{IRS}}}}} \right) - 10\alpha {\log _{10}}\left( {{x_{{\rm{UEC}}}} - {x_{{\rm{IRS}}}}} \right),
	\end{align}
	where ${x_{{\rm{UEC}}}}$ denotes the x-coordinate of the user central point.
	Thus, the minimum value of \eqref{l-s_chan_gain} is achieved at ${x^{*}_{{\rm{UEC}}}}={x_{{\rm{IRS}}}}/2$, which is consistent with the simulation results.
	Finally, as expected, the schemes that jointly optimize $\bf F$ and $\bm \phi$ significantly improve the WMR performance over the \textbf{Rand} schemes.
	The performance of the \textbf{2 bit} schemes with lower hardware cost falls in between the optimal continuous-phase and the random phase solutions, indicating that much improved performance can be obtained with even coarsely quantized phases.

	\subsection{Impact of the SI Coefficient}
	Next we focus on the effect of SI in Fig. \ref{fig:impact_loca_rho}.
	Comparing Fig. \ref{fig:xIRS-SI0.1} with Fig. \ref{fig:xIRS-SI1}, efficient user SI elimination techniques can improve the WMR when the IRS is deployed on the user side.
	Specifically, the achievable WMRs for \textbf{BCD-MM} and \textbf{SOCP+MM} schemes increase from 0.85 to 0.95 when $x_{\rm {IRS}}=120$.
	However, it should be emphasized that $\rho_{\rm S}=0.1$ is an ideal case,
	the feasibility of which needs experimental verification in real scenarios.
	It can be observed that even in this ideal scenario, the WMR achieved by deploying the IRS near the users is still lower than when the IRS is deployed near the BS.
	This is due to the fact that there is also co-channel interference (CI) in the signals received by the users.
	When the IRS is further away from the users, the impact of both SI and CI is relatively small.
	Additionally, with an increase in number of users $K$, CI gradually increases and becomes dominant.
	Thus, more of the IRS resources will be assigned to reduce the CI when it is deployed near the users.

	\subsection{Impact of Asymmetric Priority}
	\begin{figure}
		\centering
		\includegraphics[width = 1\linewidth]{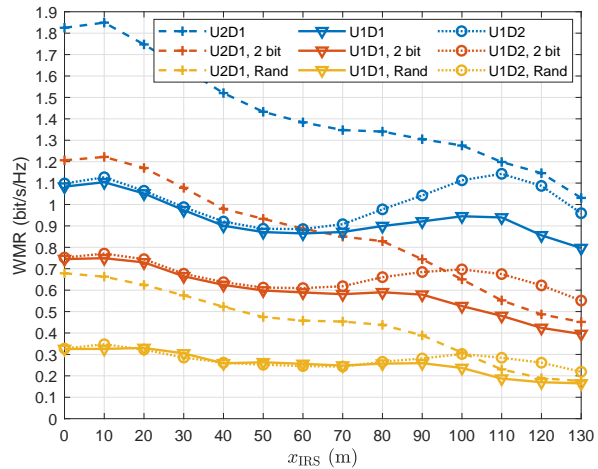}
		\caption{Impact of the IRS location $x_{\rm {IRS}}$ for asymmetric-priority scenarios.
			\label{fig:xIRS-asym}}
	\end{figure}
	As mentioned in Remark \ref{remark:weight}, the desired uplink and downlink rates in a cellular system are often asymmetric.
	To provide more comprehensive engineering guidance, in Fig. \ref{fig:xIRS-asym}, we set weighting factors $\omega_{{\rm U},k}=\omega_{{\rm U}}$ and $\omega_{{\rm D},k}=\omega_{{\rm D}}$ for $\forall k$ and study achievable WMR versus $x_{\rm {IRS}}$ in the following three scenarios various in priority condition:
	1) Downlink first ($\omega_{{\rm U}}=2, \omega_{{\rm D}}=1$, denoted as \emph{U2D1});
	2) Equal priority ($\omega_{{\rm U}}=1, \omega_{{\rm D}}=1$, denoted as \emph{U1D1});
	3) Uplink first ($\omega_{{\rm U}}=1, \omega_{{\rm D}}=2$, denoted as \emph{U1D2}).
	\textbf{BCD-MM} is  adopted in all schemes.
	It should be emphasized that the WMR, whose value is affected by the weighting factor, is not equivalent to the data rate.
	We see that for most scenarios, IRS is preferable to deployed on the BS side than on the user side.
	This contrast is obvious in the downlink first scenario, which is the most common in practice.
	A slightly different conclusion is presented only in uplink first scenarios where IRS has arbitrary phase shifts.
	Based on the discussions in this and the previous subsections, it can be concluded that in a common FD two-way communication scenario, the IRS deployment location that maximizes performance of all users is between the BS and users, near the BS.
	However, since the BS-to-user channels are blocked, moving the IRS closer to the BS may increase the likelihood that the IRS-to-user channels become blocked as well.

	\subsection{Impact of the Weights and the Achieved Fairness }
	\begin{figure}
		\centering
		\includegraphics[width = 1\linewidth]{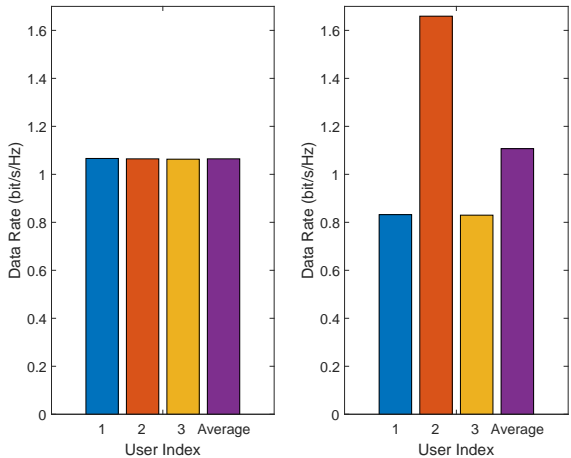}
		\caption{Individual data rate under two sets of weights.}
		\label{fig:weights}
	\end{figure}

	As mentioned in the problem formulation, the essence of guaranteeing the fairness is to allocate resources from the users with higher rates to those with lower rates, thus the data rates of all users tend to be equal.
	Additionally, the weighting factor $\omega_{l,k}$ represents the inverse of the priority of the corresponding user in the link direction $l$.
	This means that by appropriately setting $\omega_{l,k}$, multiple user characteristics can be taken into account.
	To illustrate this, we choose an example with $\omega_{{\rm D},k}=\omega_{{\rm U},k}$ for each user, and set the coordinates of the three users as $\left(100,10\right),\left(120,0\right)$ and $\left(140,-10\right)$.
	Taking the user activity levels into consideration, two scenarios are tested:
	1) Each user is active ($\omega_{k}=1,\forall k$);
	and
	2) User 2 is more active than the other two users ($\omega_{2}=1,\omega_{1}=\omega_{3}=2$).
	Fig. \ref{fig:weights} illustrates the individual data rates achieved under both scenarios.
	The average of the data rates is also plotted.
	As expected, a balanced rate distribution is obtained with equal weights, even though the path loss related to each user varies significantly.
	Additionally, the most active user 2 achieves the highest data rate in the scenario with different user activity levels.
	Furthermore, the essentially constant average rate illustrates the flexibility of the IRS-aided communication system for resource allocation.

	\subsection{Impact of the Path Loss Exponent}
	\begin{figure}
		\centering
		\includegraphics[width = 1\linewidth]{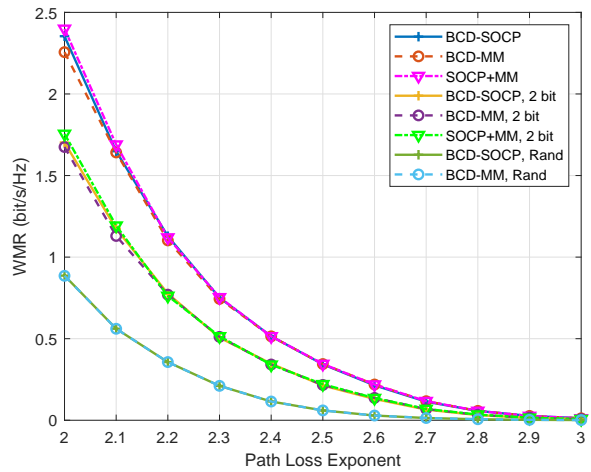}
		\caption{Achievable WMR versus path loss exponent.}
		\label{fig:PathLoss}
	\end{figure}
	In some practical scenarios, an ideal location for deploying the IRS may be infeasible, which means that path loss exponents $\alpha_{\rm {IRS}}$ as low as 2.2 may not be guaranteed.
	To investigate the system performance under different levels of fading,
	we plot Fig. \ref{fig:PathLoss} showing the achievable WMR for various path loss exponents.
	It can be observed that path loss has a significant impact on the WMR performance.
	Specifically, in each scenario, the increase in the achievable WMR is more than doubled for every 0.2 decrease in the value of $\alpha_{\rm {IRS}}$.
	Ultimately, the WMR performance decays to 0 at high values of $\alpha_{\rm {IRS}}$.
	This provides important guidance for engineering design: the performance gain obtained by deploying an IRS is greatly affected by channel conditions, thus the IRS should be deployed in a location with fewer obstacles.

	\subsection{Impact of the Rician Factor}
	\begin{figure}
		\centering
		\includegraphics[width = 1\linewidth]{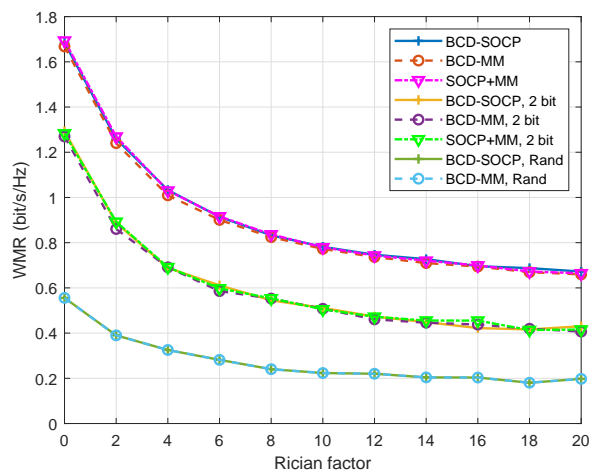}
		\caption{Achievable WMR versus the Rician factor.}
		\label{fig:Rician}
	\end{figure}
	Fig. \ref{fig:Rician} shows the achievable WMR for various Rician factors $\kappa$, which characterizes the scattering of the channel.
	As the multipath diversity gain decreases, the achievable WMR decreases as expected.
	Moreover, it can be observed that in the rich-scattering Rayleigh channel environment ($\kappa=0$),
	the achievable multipath diversity gain of the \textbf{Rand} schemes is significantly lower than that of the other methods,
	which again highlights the advantages of joint optimization.
	
	\subsection{Impact of the Number of IRS Reflection Elements}
	\begin{figure}
		\centering
		\includegraphics[width = 1\linewidth]{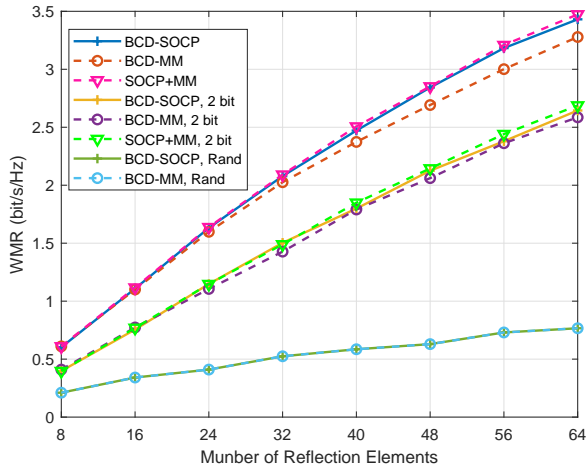}
		\caption{Achievable WMR versus the number of IRS reflection elements $M$.}
		\label{fig:M}
	\end{figure}
	According to the previous discussions, when the IRS is deployed on the BS side, doubling the number of its reflection elements $M$ can double the power of signals with little increase in interference and noise.
	Then, according to the Shannon formula, the channel capacity increases in an approximate logarithmic manner with the increase of $M$.
	Fig. \ref{fig:M} shows the achievable WMR for various values of $M$.
	As expected, the achievable WMR is not perfectly linear with the number of IRS reflection elements.
	The performance gain from increasing the elements decreases gradually.
	However, for the range of $M$ from 8 to 64, this decreasing trend is not very significant, indicating that it is cost-effective to improve communication performance by deploying more reflection elements.

	\section{Conclusions} \label{CONCLUSION}
	In this paper, we have proposed a multiuser FD two-way communication network that exploits the availability of an IRS to enhance user fairness.
	Specifically, with appropriately adjusted phase shifts, the IRS can create effective reflective paths between the BS and the users, while simultaneously mitigating the interference at the users.
	We investigated the WMR maximization problem, where the BS precoding matrix and the IRS reflection coefficients were jointly optimized subject to maximum transmit power and unit-modulus constraints.
	We transformed the original problem into an equivalent form, and then introduced the BCD algorithm to alternately optimize the variables.
	An MM algorithm with closed-form solutions in each iteration was proposed to further reduce the computational complexity.
	Our simulation results showed that the proposed algorithm has a high convergence speed in terms of both the number of iterations and CPU time, and achieves high communication performance.
	In addition, the results imply that when the IRS is deployed near the users, user SI elimination techniques can improve system performance to some extent.
	But in common scenarios, the IRS should be deployed at a location between the BS and the users with favorable reflection links, closer to the BS.
	
	\begin{appendices}

		\section{Proof of Theorem \ref{theo_of_fF}}\label{appfF}
		Note that each ${h_{{\rm{D}},k}}\left( {\bf{F}} \right),{k \in {\cal K}}$ is a quadratic function, so we propose that the minorizing function for $f\left( {\bf{F}} \right)$ has the following quadratic form:
		\begin{align}\label{tildefF_origin}
			\tilde f\left( {{\bf{F}}|{{\bf{F}}^n}} \right) &= f\left( {{{\bf{F}}^n}} \right) + 2{\mathop{\rm Re}\nolimits} \left\{ {{\mathop{\rm Tr}\nolimits} \left[ {{{\bf{D}}^{\rm H}}\left( {{\bf{F}} - {{\bf{F}}^n}} \right)} \right]} \right\} \notag \\
			&\quad + {\mathop{\rm Tr}\nolimits} \left[ {{{\left( {{\bf{F}} - {{\bf{F}}^n}} \right)}^{\rm H}}{\bf{M}}\left( {{\bf{F}} - {{\bf{F}}^n}} \right)} \right],
		\end{align}
		where ${\bf{D}}\in {\mathbb{C}}^{{N_{\rm t}} \times {N_{\rm t}}}$ and ${\bf{M}}\in {\mathbb{C}}^{{N_{\rm t}} \times {N_{\rm t}}}$ are undetermined parameters.
		Note that conditions (A1) and (A4) are already satisfied, so the expressions for $\bf D$ and $\bf M$ are determined by conditions (A2) and (A3).
		
		Let ${\bf{F}}^{\rm t}$ be a member of ${{\cal S}_{F}}$. Then, the directional derivative of $\tilde f\left( {{\bf{F}}|{{\bf{F}}^n}} \right)$ in \eqref{tildefF_origin} at ${{\bf{F}}^n}$ in direction ${\bf{F}}^{\rm t}-{{\bf{F}}^n}$ is given by
		\begin{equation}\label{direc_deriv_tildefF}
			2{\mathop{\rm Re}\nolimits} \left\{ {{\mathop{\rm Tr}\nolimits} \left[ {{{\bf{D}}^{\rm H}}\left( {{\bf{F}}^{\rm t} - {{\bf{F}}^n}} \right)} \right]} \right\}.
		\end{equation}
		In addition, the directional derivative of $f\left( {\bf{F}} \right)$ is
		\begin{equation}\label{direc_deriv_fF}
			2{\mathop{\rm Re}\nolimits} \left\{ {{\mathop{\rm Tr}\nolimits} \left[ {\sum\limits_{k \in {\cal K}} {{g_{{\rm{D}},k}}\left( {{{\bf{F}}^n}} \right)\left( {{\bf{C}}_k^{\rm H} - {{\left( {{{\bf{F}}^n}} \right)}^{\rm H}}{{\bf{B}}_k}} \right)\left( {{\bf{F}}^{\rm t} - {{\bf{F}}^n}} \right)} } \right]} \right\},
		\end{equation}
		where ${g_{D,k}}\left( {{{\bf{F}}^n}} \right)$ is defined in \eqref{gDkFn}.
		From condition (A3), the two directional derivatives \eqref{direc_deriv_tildefF} and \eqref{direc_deriv_fF} must be equal. By comparing the coefficients, the matrix ${\bf{D}}$ is identified as
		\begin{equation}
			{\bf{D}} = \sum\limits_{k \in {\cal K}} {{g_{{\rm{D}},k}}\left( {{{\bf{F}}^n}} \right)\left( {{{\bf{C}}_k} - {\bf{B}}_k^{\rm H}{{\bf{F}}^n}} \right)}.
		\end{equation}
		Then, to satisfy condition (A2), we try to make the minorizing function $\tilde f\left( {{\bf{F}}|{{\bf{F}}^n}} \right)$ be a lower bound of $f\left( {\bf{F}} \right)$ for each linear cut in any direction.
		By introducing an auxiliary variable $\eta \in \left[0,1\right]$, and letting ${\bf{F}}={{\bf{F}}^n}+\eta \left({{\bf{F}}^{\rm t}-{{\bf{F}}^n}}\right)$, this sufficient condition can be expressed as
		\begin{align}\label{sufficient_A2}
			&f\left({{\bf{F}}^n}+\eta \left({{\bf{F}}^{\rm t}-{{\bf{F}}^n}}\right)\right) \notag\\
			&\qquad\geq f\left({\bf{F}}^n\right)+2\eta{\mathop{\rm Re}\nolimits}\left\{ {{\mathop{\rm Tr}\nolimits} \left[ {{{\bf{D}}^{\rm H}}\left( {{\bf{F}}^{\rm t} - {{\bf{F}}^n}} \right)} \right]} \right\}\notag\\
			&\qquad\quad+\eta^2{\mathop{\rm Tr}\nolimits} \left[ {{{\left( {{\bf{F}}^{\rm t} - {{\bf{F}}^n}} \right)}^{\rm H}}{\bf{M}}\left( {{\bf{F}}^{\rm t} - {{\bf{F}}^n}} \right)} \right].
		\end{align}
		
		Denote the left and right hand side of \eqref{sufficient_A2} by ${j_F}\left( \eta  \right)$ and ${J_F}\left( \eta  \right)$, respectively.
		Then, it is apparent that ${j_F}\left( 0  \right)={J_F}\left( 0  \right)$.
		The first-order derivative of ${j_F}\left( \eta  \right)$ is calculated as
		\begin{equation}\label{first_deriv_j}
			{\nabla _\eta }{j_F}\left( \eta  \right) = \sum\limits_{k \in {\cal K}} {{{\hat{g}}_{{\rm{D}},k}}\left( \eta  \right){\nabla _\eta }{{\hat{h}}_{{\rm{D}},k}}\left( \eta  \right)},
		\end{equation}
		where
		\begin{subequations}
			\begin{alignat}{2}
				&{{\hat{h}}_{{\rm{D}},k}}\left( \eta  \right)\buildrel \Delta \over={{{h}}_{{\rm{D}},k}} {\left({{\bf{F}}^n}+\eta \left({{\bf{F}}^{\rm t}-{{\bf{F}}^n}}\right)\right)},\\
				&{{\hat{g}}_{{\rm{D}},k}}\left( \eta  \right)\buildrel \Delta \over=\frac{{\exp \left\{ { - \mu {{\hat{h}}_{{\rm{D}},k}}\left( {\eta} \right)} \right\}}}{{\sum\limits_{k \in {\cal K}} {\exp \left\{ { - \mu {{\hat{h}}_{{\rm{D}},k}}\left( {\eta} \right)} \right\}} }},k \in {\cal K}.
			\end{alignat}
		\end{subequations}
		And it can be derived that
		\begin{align}\label{deriv_hat_hDk}
			&{\nabla _\eta }{\hat h_{{\rm{D}},k}}\left( \eta  \right) \nonumber\\
			&= 2{\rm{Re}}\left\{ {{\rm{Tr}}\left( {{\bf{C}}_k^{\rm{H}}\left( {{{\bf{F}}^{\rm{t}}} - {{\bf{F}}^n}} \right) - {{\left( {{{\bf{F}}^n}} \right)}^{\rm{H}}}{{\bf{B}}_k}\left( {{{\bf{F}}^{\rm{t}}} - {{\bf{F}}^n}} \right)} \right)} \right\} \nonumber\\
			&\quad - 2\eta {\rm{Tr}}\left( {{{\left( {{{\bf{F}}^{\rm{t}}} - {{\bf{F}}^n}} \right)}^{\rm{H}}}{{\bf{B}}_k}\left( {{{\bf{F}}^{\rm{t}}} - {{\bf{F}}^n}} \right)} \right).
		\end{align}
		It is readily verified that ${\nabla _\eta }{j_F}\left( 0\right)={\nabla _\eta }{J_F}\left( 0 \right)$.
		Then, since ${J_F}\left( \eta  \right)$ is concave w.r.t. $\eta$, a sufficient condition for \eqref{sufficient_A2} to hold is that the second-order derivative of ${j_F}\left( \eta  \right)$ is greater than or equal to that of ${J_F}\left( \eta  \right)$ for $\forall \eta \in \left[0,1\right]$, i.e.
		\begin{equation}\label{2nd_deriv_j_geq_J_F}
			\nabla _\eta ^2 {j_F}\left( \eta  \right) \geq \nabla _\eta ^2 {J_F}\left( \eta  \right), \forall \eta \in \left[0,1\right].
		\end{equation}
		
		In the following, we compute the second-order derivative of ${j_F}\left( \eta  \right)$ to determine the value of ${\bf{M}}$.
		First, by defining
		\[{\bf{E}}_k  \buildrel \Delta \over =  {{\bf{C}}_k} - {\bf{B}}_k^{\rm H}\left( {{{\bf{F}}^n} + \eta \left( {{{\bf{F}}^t} - {{\bf{F}}^n}} \right)} \right),\]
		\eqref{deriv_hat_hDk} can be rewritten as
		\begin{align}
			{\nabla _\eta }{\hat h_{{\rm{D}},k}}\left( \eta  \right) &=2{\rm{Re}}\left\{ {{\rm{Tr}}\left( {{\bf{E}}_k^{\rm{H}}\left( {{{\bf{F}}^t} - {{\bf{F}}^n}} \right)} \right)} \right\}\notag \\
			&=2{\mathop{\rm Re}\nolimits} \left\{ {{\bf{e}}_k^{\rm H}{\bf{\bar f}}} \right\},
		\end{align}
		where ${{\bf{e}}_k} \buildrel \Delta \over = {\mathop{\rm vec}\nolimits} \left( {{{\bf{E}}_k}} \right)$ and ${{\bf{\bar f}}} \buildrel \Delta \over = {\mathop{\rm vec}\nolimits} \left( {{{\bf{F}}^t} - {{\bf{F}}^n}} \right)$.
		The second-order derivative of ${\hat h_{{\rm{D}},k}}\left( \eta  \right)$ is given by
		\begin{align}
			\nabla _\eta ^2{\hat h_{D,k}}\left( \eta  \right) &=  - 2{\mathop{\rm Tr}\nolimits} \left( {{{\left( {{{\bf{F}}^t} - {{\bf{F}}^n}} \right)}^{\rm{H}}}{{\bf{B}}_k}\left( {{{\bf{F}}^t} - {{\bf{F}}^n}} \right)} \right)\notag \\
			&=- 2{{\bf{\bar f}}^{\rm H}}\left( {{\bf{I}} \otimes {{\bf{B}}_k}} \right){\bf{\bar f}},
		\end{align}
		where we have used the property that ${\mathop{\rm Tr}\nolimits} \left( {{\bf{ABC}}} \right) = {{\mathop{\rm vec}\nolimits} ^{\rm T}}\left( {{{\bf{A}}^{\rm T}}} \right)\left( {{\bf{I}} \otimes {\bf{B}}} \right){\mathop{\rm vec}\nolimits} \left( {\bf{C}} \right)$ \cite{Zhang2017Matrix}.
		Then, the second-order derivative of ${j_F}\left( \eta  \right)$ is derived as
		\begin{align}
			&\nabla _\eta ^2 {j_F}\left( \eta  \right) \notag \\
			&=  \sum\limits_{k \in {\cal K}} {\left( {{{\hat g}_{{\rm{D}},k}}\left( \eta  \right)\nabla _\eta ^2{{\hat h}_{{\rm{D}},k}}\left( \eta  \right) - \mu {{\hat g}_{{\rm{D}},k}}\left( \eta  \right){{\left( {{\nabla _\eta }{{\hat h}_{{\rm{D}},k}}\left( \eta  \right)} \right)}^2}} \right)} \quad \notag \\
			&\quad + \mu {\left( {\sum\limits_{k \in {\cal K}} {{{\hat g}_{{\rm{D}},k}}\left( \eta  \right){\nabla _\eta }{{\hat h}_{{\rm{D}},k}}\left( \eta  \right)} } \right)^2} \notag \\
			&= \left[ {\begin{array}{*{20}{c}}
					{{{{\bf{\bar f}}}^{\rm H}}}&{{{{\bf{\bar f}}}^{\rm T}}}
			\end{array}} \right]\Xi \left[ {\begin{array}{*{20}{c}}
					{{\bf{\bar f}}}\\
					{{{{\bf{\bar f}}}^ * }}
			\end{array}} \right],
		\end{align}
		where $\Xi$ is given by \eqref{Xi} on the top of the next page.
		\begin{figure*}[ht]
			\begin{equation}\label{Xi}
				{\bf{\Xi }}  =  - \sum\limits_{k \in {\cal K}} {{{\hat g}_{{\rm{D}},k}}\left( \eta  \right)\left( {\left[ {\begin{array}{*{20}{c}}
								{{\bf{I}} \otimes {{\bf{B}}_k}}&{\bf{0}}\\
								{\bf{0}}&{{\bf{I}} \otimes {\bf{B}}_k^{\rm H}}
						\end{array}} \right] + \mu \left[ {\begin{array}{*{20}{c}}
								{{{\bf{e}}_k}}\\
								{{\bf{e}}_k^ * }
						\end{array}} \right]{{\left[ {\begin{array}{*{20}{c}}
										{{{\bf{e}}_k}}\\
										{{\bf{e}}_k^ * }
								\end{array}} \right]}^{\rm H}}} \right)}  + \mu \left[ {\begin{array}{*{20}{c}}
						{\sum\limits_{k \in {\cal K}} {{{\hat g}_{D,k}}\left( \eta  \right){{\bf{e}}_k}} }\\
						{\sum\limits_{k \in {\cal K}} {{{\hat g}_{D,k}}\left( \eta  \right){\bf{e}}_k^ * } }
				\end{array}} \right]{\left[ {\begin{array}{*{20}{c}}
							{\sum\limits_{k \in {\cal K}} {{{\hat g}_{D,k}}\left( \eta  \right){{\bf{e}}_k}} }\\
							{\sum\limits_{k \in {\cal K}} {{{\hat g}_{D,k}}\left( \eta  \right){\bf{e}}_k^ * } }
					\end{array}} \right]^{\rm H}}.
			\end{equation}
			\hrulefill
			\vspace{-0.3cm}
		\end{figure*}
		
		We also compute the second-order derivative of $\nabla _\eta ^2 {J_F}\left( \eta  \right)$, and manipulate it into a quadratic form, as follows
		\begin{align}
			\nabla _\eta ^2 {J_F}\left( \eta  \right) &= 2{\mathop{\rm Tr}\nolimits} \left[ {{{\left( {{\bf{F}}^{\rm t} - {{\bf{F}}^n}} \right)}^{\rm H}}{\bf{M}}\left( {{\bf{F}}^{\rm t} - {{\bf{F}}^n}} \right)} \right] \notag \\
			& =\left[ {\begin{array}{*{20}{c}}
					{{{{\bf{\bar f}}}^{\rm H}}}&{{{{\bf{\bar f}}}^{\rm T}}}
			\end{array}} \right]\left[ {\begin{array}{*{20}{c}}
					{{\bf{I}} \otimes {\bf{M}}}&{\bf{0}}\\
					{\bf{0}}&{{\bf{I}} \otimes {{\bf{M}}^{\rm T}}}
			\end{array}} \right] \left[ {\begin{array}{*{20}{c}}
					{{\bf{\bar f}}}\\
					{{{{\bf{\bar f}}}^ * }}
			\end{array}} \right].
		\end{align}
		Then, the inequality in \eqref{2nd_deriv_j_geq_J_F} is reformulated as
		\begin{align}
			&\left[ {\begin{array}{*{20}{c}}
					{{{{\bf{\bar f}}}^{\rm H}}}&{{{{\bf{\bar f}}}^{\rm T}}}
			\end{array}} \right]{\bf{\Xi }} \left[ {\begin{array}{*{20}{c}}
					{{\bf{\bar f}}}\\
					{{{{\bf{\bar f}}}^ * }}
			\end{array}} \right] \notag\\
			&\qquad\geq \left[ {\begin{array}{*{20}{c}}
					{{{{\bf{\bar f}}}^{\rm H}}}&{{{{\bf{\bar f}}}^{\rm T}}}
			\end{array}} \right]\left[ {\begin{array}{*{20}{c}}
					{{\bf{I}} \otimes {\bf{M}}}&{\bf{0}}\\
					{\bf{0}}&{{\bf{I}} \otimes {{\bf{M}}^{\rm T}}}
			\end{array}} \right] \left[ {\begin{array}{*{20}{c}}
					{{\bf{\bar f}}}\\
					{{{{\bf{\bar f}}}^ * }}
			\end{array}} \right].
		\end{align}
		As a result, ${\bf{M}}$ must satisfy
		\begin{equation}
			{\bf{\Xi }} \succeq \left[ {\begin{array}{*{20}{c}}
					{{\bf{I}} \otimes {\bf{M}}}&{\bf{0}}\\
					{\bf{0}}&{{\bf{I}} \otimes {{\bf{M}}^{\rm T}}}
			\end{array}} \right].
		\end{equation}
		
		We choose the following simple solution: ${\bf{M}}=\alpha {\bf{I}}={\lambda _{\min }}\left({\bf{\Xi }}\right){\bf{I}}$.
		Then, \eqref{tildefF_origin} is equivalent to
		\begin{align}\label{tildefF_endofAppA}
			\tilde f\left( {{\bf{F}}|{{\bf{F}}^n}} \right) &= f\left( {{{\bf{F}}^n}} \right) + 2{\mathop{\rm Re}\nolimits} \left\{ {{\mathop{\rm Tr}\nolimits} \left[ {{{\bf{D}}^{\rm H}}\left( {{\bf{F}} - {{\bf{F}}^n}} \right)} \right]} \right\} \notag  \\
			&\quad + \alpha{\mathop{\rm Tr}\nolimits} \left[ {{{\left( {{\bf{F}} - {{\bf{F}}^n}} \right)}^{\rm H}}\left( {{\bf{F}} - {{\bf{F}}^n}} \right)} \right] \notag \\
			&={2{\mathop{\rm Re}\nolimits} \left\{ {{\mathop{\rm Tr}\nolimits} \left[ {{{\bf{V}}^{\rm H}}{\bf{F}}} \right]} \right\} + \alpha {\mathop{\rm Tr}\nolimits} \left[ {{{\bf{F}}^{\rm H}}{\bf{F}}} \right] + {\rm{cons}}F},
		\end{align}
		where ${\bf{V}}$ and ${\rm{cons}}F$ are given in \eqref{VarforF_V} and \eqref{tildefF_consF}, respectively.
		However, ${\bf{\Xi }}$ is a very complex function w.r.t. $\eta$, which leads to a high computation cost to calculate $\alpha$ in \eqref{tildefF_endofAppA}.
		To reduce the complexity, we proceed to find a simple lower bound to replace $\alpha$, as shown in \eqref{alpha_lowbound} on the top of the next page,
		\begin{figure*}[ht]
			\begin{align}\label{alpha_lowbound}
				\alpha= {\lambda _{\min }}\left( {\bf{\Xi }} \right) &\overset{(a1)}{\ge}  - \sum\limits_{k \in {\cal K}} {{{\hat g}_{{\rm{D}},k}}\left( \eta  \right)\left( {{\lambda _{\max }}\left( {\left[ {\begin{array}{*{20}{c}}
									{{\bf{I}} \otimes {{\bf{B}}_k}}&{\bf{0}}\\
									{\bf{0}}&{{\bf{I}} \otimes {\bf{B}}_k^{\rm{H}}}
							\end{array}} \right]} \right) + \mu {\lambda _{\max }}\left( {\left[ {\begin{array}{*{20}{c}}
									{{{\bf{e}}_k}}\\
									{{\bf{e}}_k^*}
							\end{array}} \right]{{\left[ {\begin{array}{*{20}{c}}
											{{{\bf{e}}_k}}\\
											{{\bf{e}}_k^*}
									\end{array}} \right]}^{\rm{H}}}} \right)} \right)} \notag \\
				&\qquad + \mu {\lambda _{\min }}\left( {\left[ {\begin{array}{*{20}{c}}
							{\sum\limits_{k \in {\cal K}} {{{\hat g}_{{\rm{D}},k}}\left( \eta  \right){{\bf{e}}_k}} }\\
							{\sum\limits_{k \in {\cal K}} {{{\hat g}_{{\rm{D}},k}}\left( \eta  \right){\bf{e}}_k^*} }
					\end{array}} \right]{{\left[ {\begin{array}{*{20}{c}}
									{\sum\limits_{k \in {\cal K}} {{{\hat g}_{{\rm{D}},k}}\left( \eta  \right){{\bf{e}}_k}} }\\
									{\sum\limits_{k \in {\cal K}} {{{\hat g}_{{\rm{D}},k}}\left( \eta  \right){\bf{e}}_k^*} }
							\end{array}} \right]}^{\rm{H}}}} \right)\notag\\
				&\overset{(a2)}{=} - \sum\limits_{k \in {\cal K}} {{{\hat g}_{{\rm{D}},k}}\left( \eta  \right)\left( {{\lambda _{\max }}\left( {{{\bf{B}}_k}} \right) + 2\mu {\bf{e}}_k^{\rm H}{{\bf{e}}_k}} \right)}\notag\\
				&\overset{(a2)}= - \sum\limits_{k \in {\cal K}} {{{\hat g}_{{\rm{D}},k}}\left( \eta  \right)\left( {\rm{tp1}}_{k} \right)}  - 2\mu \sum\limits_{k \in {\cal K}} {{{\hat g}_{{\rm{D}},k}}\left( \eta  \right)\left\| {{{\bf{E}}_k}} \right\|_F^2}\notag\\
				&\overset{(a3)}\geq - \mathop {\max }\limits_{k} \left\{ {\rm{tp1}}_{k} \right\} - 2\mu \mathop {\max }\limits_{k } \left\{ {\left\| {{{\bf{E}}_k}} \right\|_F^2} \right\},
			\end{align}
			\hrulefill
		\end{figure*}
		where ${{\rm {tp1}}_k}$ is defined in \eqref{tp1k}, and we have used the following properties (a1)-(a3):
		\begin{enumerate}[({a}1)]
			\setlength{\itemsep}{0.7ex}
			\item ${\lambda _{\min }}\left( {\bf{A}} \right) + {\lambda _{\min }}\left( {\bf{B}} \right) \le {\lambda _{\min }}\left( {{\bf{A}} + {\bf{B}}} \right)$, if ${\bf{A}}$ and ${\bf{B}}$ are Hermitian matrices \cite{lutkepohl1996handbook};
			
			\item ${\lambda _{\max }}\left( {\bf{A}} \right) = {\mathop{\rm Tr}\nolimits} \left( {\bf{A}} \right)$ and ${\lambda _{\min }}\left( {\bf{A}} \right) = 0$, if ${\bf{A}}$ is rank one \cite{lutkepohl1996handbook};
			
			\item $\sum\nolimits_{m = 1}^M {{a_m}{b_m}}  \le \max _{m = 1}^M\left\{ {{b_m}} \right\}$, if ${a_m},{b_m}\ge 0$ and $\sum\nolimits_{m = 1}^M {{a_m}}=1$ \cite[Theorem 30]{Matrix-differential}.
			
		\end{enumerate}
		
		Recall that ${\bf{F}}={{\bf{F}}^n}+\eta \left({{\bf{F}}^t-{{\bf{F}}^n}}\right)$,
		thus the inequality $\left\| {{{\bf{F}}^n} + \eta \left( {{{\bf{F}}^t} - {{\bf{F}}^n}} \right)} \right\|_F \le \sqrt{P_{\max }}$ holds.
		Then an upper bound for $\left\| {{{\bf{E}}_k}} \right\|_F^2$ is derived in \eqref{EkF2_lowbound} on the top of the next page,
		\begin{figure*}[ht]
			\begin{align}\label{EkF2_lowbound}
				\left\| {{{\bf{E}}_k}} \right\|_F^2 &= \left\| {{{\bf{C}}_k} - {\bf{B}}_k^{\rm{H}}\left( {{{\bf{F}}^n} + \eta \left( {{{\bf{F}}^t} - {{\bf{F}}^n}} \right)} \right)} \right\|_F^2\notag\\
				&=\left\| {{\bf{B}}_k^{\rm{H}}\left( {{{\bf{F}}^n} + \eta \left( {{{\bf{F}}^t} - {{\bf{F}}^n}} \right)} \right)} \right\|_F^2 + \left\| {{{\bf{C}}_k}} \right\|_F^2 - 2{\mathop{\rm Re}\nolimits} \left\{ {{\mathop{\rm Tr}\nolimits} \left( {{\bf{C}}_k^{\rm H}{\bf{B}}_k^{\rm{H}}\left( {{{\bf{F}}^n} + \eta \left( {{{\bf{F}}^t} - {{\bf{F}}^n}} \right)} \right)} \right)} \right\}\notag\\
				&\overset{(a4)}{\leq} {\lambda _{\max }}\left( {{\bf{B}}_k^{\rm{H}}{{\bf{B}}_k}} \right)\left\| {{{\bf{F}}^n} + \eta \left( {{{\bf{F}}^t} - {{\bf{F}}^n}} \right)} \right\|_F^2 + \left\| {{{\bf{C}}_k}} \right\|_F^2 - 2{\mathop{\rm Re}\nolimits} \left\{ {{\mathop{\rm Tr}\nolimits} \left( {{\bf{C}}_k^{\rm H}{\bf{B}}_k^{\rm{H}}\left( {{{\bf{F}}^n} + \eta \left( {{{\bf{F}}^t} - {{\bf{F}}^n}} \right)} \right)} \right)} \right\}\notag\\
				& \overset{(a5)} \le {P_{\max }}{\lambda _{\max }}\left( {{\bf{B}}_k^{\rm{H}}{{\bf{B}}_k}} \right) + \left\| {{{\bf{C}}_k}} \right\|_F^2 + 2\sqrt {{P_{\max }}} {\left\| {{{\bf{B}}_k}{{\bf{C}}_k}} \right\|_F}\notag\\
				&\overset{(a2)}={P_{\max }}{\rm{tp1}}_{k}^2 + \left\| {{{\bf{C}}_k}} \right\|_F^2 + 2\sqrt {{P_{\max }}} {\left\| {{{\bf{B}}_k}{{\bf{C}}_k}} \right\|_F}.
			\end{align}
			\hrulefill
		\end{figure*}
		where (a4) and (a5) are given by
		\begin{enumerate}[({a}1)]\setcounter{enumi}{3}
			\setlength{\itemsep}{0.7ex}
			\item ${\mathop{\rm Tr}\nolimits} \left( {{\bf{AB}}} \right) \le {\lambda _{\max }}\left( {\bf{A}} \right){\mathop{\rm Tr}\nolimits} \left( {\bf{B}} \right)$, if $\bf A$ and $\bf B$ are positive semidefinite matrices \cite{lutkepohl1996handbook};
			
			\item $ - \sqrt {{P_{\max }}} {\left\| {{{\bf{B}}}{{\bf{C}}}} \right\|_F}$ is the optimal value of the following Problem \eqref{problem_in_a5}:
			\begin{subequations}\label{problem_in_a5}
				\begin{alignat}{2}
					\min_{\bf{X}} \quad & {\mathop{\rm Re}\nolimits} \left\{ {{\mathop{\rm Tr}\nolimits} \left( {{{\bf{C}}^{\rm H}}{{\bf{B}}^{\rm H}}{\bf{X}}} \right)} \right\} \\
					\mbox{s.t.}\quad
					&{\mathop{\rm Tr}\nolimits} \left( {{{\bf{X}}^{\rm H}}{\bf{X}}} \right) \le {P_{\max }}.
				\end{alignat}
			\end{subequations}
			
		\end{enumerate}
		Finally, by substituting \eqref{EkF2_lowbound} into \eqref{alpha_lowbound}, we arrive at \eqref{tildefF}.
		Hence, the proof is complete.
		
		\section{Proof of Theorem \ref{theo_of_fphi}}\label{appfphi}
		We propose a quadratic function to minorize $f\left({\bm{\phi}}\right)$.
		Defining undetermined parameters ${\bf{N}}\in {\mathbb{C}}^{{M} \times {M}}$ and ${\bf{d}}\in {\mathbb{C}}^{{M} \times {1}}$, the minorizing function $\tilde f\left( {{\bm{\phi}} |{{\bm{\phi}} ^n}} \right)$ can be expressed as
		\begin{align}\label{tildefphi_origin}
			\tilde f\left( {{\bm{\phi}} |{{\bm{\phi}} ^n}} \right) &= f\left( {{{\bm{\phi}} ^n}} \right) + 2{\rm{Re}}\left\{ {{{\bf{d}}^{\rm{H}}}\left( {{\bm{\phi}}  - {{\bm{\phi}} ^n}} \right)} \right\}\notag \\
			&\quad  + {\left( {{\bm{\phi}}  - {{\bm{\phi}} ^n}} \right)^{\rm{H}}}{\bf{N}}\left( {{\bm{\phi}}  - {{\bm{\phi}} ^n}} \right).
		\end{align}
		Since conditions (B1) and (B4) are already satisfied, in the following, we determine expressions for ${\bf{N}}$ and ${\bf{d}}$ to satisfy (B2) and (B3).
		
		Beginning with (B3), the directional derivative of $\tilde f\left( {{\bm{\phi}} |{{\bm{\phi}} ^n}} \right)$ at ${{\bm{\phi}} ^n}$ with direction $\left({\bm{\phi}}^{\rm t}-{\bm{\phi}}^n\right)$ is
		\begin{equation}\label{direc_deriv_tildefphi}
			2{\rm{Re}}\left\{ {{{\bf{d}}^{\rm{H}}}\left( {{\bm{\phi}}^{\rm t}  - {{\bm{\phi}} ^n}} \right)} \right\},
		\end{equation}
		where ${\bm{\phi}}^{\rm t} \in {\cal S}_{\phi}$.
		Applying (B3), the directional derivative of $f\left( {\bm{\phi}} \right)$ must be equal to the directional derivative \eqref{direc_deriv_tildefphi}, which means
		\begin{equation}
			{\bf{d}} = \sum\limits_{l \in {\cal L}} {\sum\limits_{k \in {\cal K}} {{g_{l,k}}\left( {\bm{\phi}}^n  \right)\left( {{{\bf{a}}_{l,k}} - {\bf{A}}_{l,k}^{\rm H}{{\bm{\phi}} ^n}} \right)} },
		\end{equation}
		where ${g_{l,k}}\left( {\bm{\phi}}^n  \right)$ is defined in \eqref{glkphin}.
		
		Now we consider condition (B2).
		Let ${\bm{\phi}}={\bm{\phi}}^n+\eta\left({\bm{\phi}}^{\rm t}-{\bm{\phi}}^n\right)$ with $\eta \in \left[0,1\right]$.
		Then a sufficient condition for (B2) is given by
		\begin{align}\label{sufficient_B2}
			f\left( {{{\bm{\phi}}^n} + \eta \left( {{{\bm{\phi}}^{\rm t}} - {{\bm{\phi}}^n}} \right)} \right) &\ge f\left( {{{\bm{\phi}}^n}} \right) + 2\eta {\rm{Re}}\left\{{{{\bf{d}}^{\rm{H}}}\left( {{{\bm{\phi}}^{\rm t}} - {{\bm{\phi}}^n}} \right)} \right\} \notag \\
			&\quad + {\eta ^2}{\left( {{{\bm{\phi}}^{\rm t}} - {{\bm{\phi}}^n}} \right)^{\rm{H}}}{\bf{N}}\left( {{{\bm{\phi}}^{\rm t}} - {{\bm{\phi}}^n}} \right).
		\end{align}
		Denote the left and right hand side of \eqref{sufficient_B2} by ${j_{\phi}}\left( \eta  \right)$ and ${J_{\phi}}\left( \eta  \right)$, respectively.
		Then we have ${j_{\phi}}\left( 0\right)={J_{\phi}}\left( 0 \right)$ and ${\nabla _\eta }{j_{\phi}}\left( 0\right)={\nabla _\eta }{J_{\phi}}\left( 0 \right)$.
		Since ${J_{\phi}}$ is concave w.r.t. $\eta$, a sufficient condition for \eqref{sufficient_B2} to hold is
		\begin{equation}\label{2nd_deriv_j_geq_J_phi}
			\nabla _\eta ^2 {j_\phi}\left( \eta  \right) \geq \nabla _\eta ^2 {J_{\phi}}\left( \eta  \right).
		\end{equation}
		With the definition ${\bar{\bm{\phi}}}\buildrel \Delta \over={{{\bm{\phi}}^{\rm t}} - {{\bm{\phi}}^n}}$, the second-order derivative of ${j_{\phi}}\left( \eta  \right)$ is given by
		\begin{equation}\label{2nd_deriv_j_phi}
			\nabla _\eta ^2 {j_\phi}\left( \eta  \right)=\left[ {\begin{array}{*{20}{c}}
					{{\bar{\bm{\phi}} ^{\rm H}}}&{{\bar{\bm{\phi}} ^{\rm T}}}
			\end{array}} \right]{\bf \Omega} \left[ {\begin{array}{*{20}{c}}
					\bar{\bm{\phi}} \\
					{{\bar{\bm{\phi}} ^ * }}
			\end{array}} \right],
		\end{equation}
		where ${\bf{\Omega }}$ is given in \eqref{appOmega} on the top of the next page.
		\begin{figure*}[ht]
			\begin{equation} \label{appOmega}
				\begin{split}
					&{\bf{\Omega }}=  - \sum\limits_{l \in {\cal L}} {\sum\limits_{k \in {\cal K}} {{{\hat{g}}_{l,k}}\left( \eta  \right)\left( {\left[ {\begin{array}{*{20}{c}}
										{{{\bf{A}}_{l,k}}}&{\bf{0}}\\
										{\bf{0}}&{{\bf{A}}_{l,k}^{\rm T}}
								\end{array}} \right] +\mu \left[ {\begin{array}{*{20}{c}}
										{{{\bf{u}}_{l,k}}}\\
										{{\bf{u}}_{l,k}^ * }
								\end{array}} \right]{{\left[ {\begin{array}{*{20}{c}}
												{{{\bf{u}}_{l,k}}}\\
												{{\bf{u}}_{l,k}^ * }
										\end{array}} \right]}^{\rm H}}} \right)} }   + \mu \left[ {\begin{array}{*{20}{c}}
							{\sum\limits_{l \in {\cal L}} {\sum\limits_{k \in {\cal K}} {{{\hat{g}}_{l,k}}\left( \eta  \right){{\bf{u}}_{l,k}}} } }\\
							{\sum\limits_{l \in {\cal L}} {\sum\limits_{k \in {\cal K}} {{{\hat{g}}_{l,k}}\left( \eta  \right){\bf{u}}_{l,k}^ * } } }
					\end{array}} \right]{\left[ {\begin{array}{*{20}{c}}
								{\sum\limits_{l \in {\cal L}} {\sum\limits_{k \in {\cal K}} {{{\hat{g}}_{l,k}}\left( \eta  \right){{\bf{u}}_{l,k}}} } }\\
								{\sum\limits_{l \in {\cal L}} {\sum\limits_{k \in {\cal K}} {{{\hat{g}}_{l,k}}\left( \eta  \right){\bf{u}}_{l,k}^ * } } }
						\end{array}} \right]^{\rm H}}.
				\end{split}
			\end{equation}
			\hrulefill
		\end{figure*}
		In \eqref{appOmega}, ${{\hat{g}}_{l,k}}$ is defined as
		\begin{equation}
			{{\hat{g}}_{l,k}}\left( \eta  \right)\buildrel \Delta \over=\frac{{\exp \left\{ { - \mu {{\hat{h}}_{l,k}}\left( {\eta} \right)} \right\}}}{\sum\limits_{l \in {\cal L}}{\sum\limits_{k \in {\cal K}} {\exp \left\{ { - \mu {{\hat{h}}_{l,k}}\left( {\eta} \right)} \right\}} }},
		\end{equation}
		where
		\begin{equation}
			{{\hat{h}}_{l,k}}\left( \eta  \right)\buildrel \Delta \over={{{h}}_{l,k}} {\left( {{{\bm{\phi}}^n} + \eta \left( {{{\bm{\phi}}^{\rm t}} - {{\bm{\phi}}^n}} \right)} \right)}.
		\end{equation}
		The second-order derivative of ${J_{\phi}}\left( \eta  \right)$ is
		\begin{equation}\label{2nd_deriv_J_phi}
			\begin{split}
				\nabla _\eta ^2 {J_{\phi}}\left( \eta  \right)&=\left[ {\begin{array}{*{20}{c}}
						{{\bar{\bm{\phi}} ^{\rm H}}}&{{\bar{\bm{\phi}} ^{\rm T}}}
				\end{array}} \right]\left[ {\begin{array}{*{20}{c}}
						{{\bf{I}} \otimes {\bf{N}}}&{\bf{0}}\\
						{\bf{0}}&{{\bf{I}} \otimes {{\bf{N}}^{\rm T}}}
				\end{array}} \right] \left[ {\begin{array}{*{20}{c}}
						\bar{\bm{\phi}} \\
						{{\bar{\bm{\phi}} ^ * }}
				\end{array}} \right].
			\end{split}
		\end{equation}

		Substituting the second-order derivatives \eqref{2nd_deriv_j_phi} and \eqref{2nd_deriv_J_phi} into \eqref{2nd_deriv_j_geq_J_phi}, we have
		\begin{equation}
			{\bf \Omega} \succeq\left[ {\begin{array}{*{20}{c}}
					{{\bf{I}} \otimes {\bf{N}}}&{\bf{0}}\\
					{\bf{0}}&{{\bf{I}} \otimes {{\bf{N}}^{\rm T}}}
			\end{array}} \right].
		\end{equation}
		For simplicity, we choose ${\bf{N}} = \beta {\bf{I}} = {\lambda _{\min }}\left( {\bf{\Omega }} \right){\bf{I}}$.
		In order to reduce the algorithm complexity, we replace $\beta$ with its lower bound, which is shown in \eqref{Varforphi_beta}.
		The method to obtain the lower bound for $\beta$ is similar as that for $\alpha$, so we omit it here.
		
		Finally, from the unit-modulus constraints on $\bm{\phi}$, we have ${\bm{\phi} ^{\rm H}}\bm{\phi}  = {\left( {{\bm{\phi} ^n}} \right)^{\rm H}}\left( {{\bm{\phi} ^n}} \right) = M$.
		Thus, \eqref{tildefphi_origin} is derived as
		\begin{align}
			\tilde f\left( {{\bm{\phi}} | {{\bm{\phi}} ^n}} \right) &= f\left( {{{\bm{\phi}} ^n}} \right) + 2{\rm{Re}}\left\{ {{{\bf{d}}^{\rm{H}}}\left( {{\bm{\phi}}  - {{\bm{\phi}} ^n}} \right)} \right\} \notag \\
			&\quad + \beta{\left( {{\bm{\phi}}  -{{\bm{\phi}} ^n}} \right)^{\rm{H}}}\left( {{\bm{\phi}}  - {{\bm{\phi}} ^n}} \right)\notag \\
			&\quad =2{\mathop{\rm Re}\nolimits} \left\{ {{{\bf{v}}^{\rm H}}{\bm{\phi}} } \right\} + {\rm{cons}}\phi,
		\end{align}
		where $\bf v$ and ${\rm{cons}}\phi$ are given in \eqref{Varforphi_v} and \eqref{Varforphi_consphi}, respectively.
		Hence, the proof is completed.
		
	\end{appendices}
	
	\bibliographystyle{IEEEtran}
	\bibliography{IEEEabrv,Refer}
	
	%

\end{document}